\documentclass[12pt, oneside]{article}
\usepackage{amsmath, amssymb}
\usepackage{dsfont}
\usepackage[english]{babel}
\usepackage{verbatim}
\usepackage{subfigure}
\usepackage{caption}
\usepackage{float}
\usepackage{url}
\usepackage{appendix}
\usepackage[square, comma, sort&compress, numbers]{natbib}
\usepackage{graphicx}

\def\bi{\begin{itemize}}     \def\ei{\end{itemize}}
\def\ben{\begin{enumerate}}   \def\een{\end{enumerate}}
\def\beq{\begin{equation}}   \def\eeq{\end{equation}}
\def\bd{\begin{displaymath}} \def\ed{\end{displaymath}}
\def\bea{\begin{eqnarray}}   \def\beaz{\begin{eqnarray*}}
\def\eea{\end{eqnarray}}     \def\eeaz{\end{eqnarray*}}
\def\beac{\begin{eqnarrayc}} \def\eeac{\end{eqnarrayc}}

\newcommand{\bc}{\begin{center}}
\newcommand{\ec}{\end{center}}
\newcommand{\be}{\begin{equation}}
\newcommand{\ee}{\end{equation}}
\newcommand{\bq}{\begin{eqnarray}}
\newcommand{\eq}{\end{eqnarray}}

\newcommand{\fp}[2]{\frac{\partial #1}{\partial #2}}

\newcommand{\nfp}[3]{\frac{\partial^{#1} #2}{\partial #3^{#1}}}

\renewcommand{\eqref}[1]{Eq.(\ref{#1})}

\begin{document}

\title{Pricing options on illiquid assets with liquid proxies using utility indifference and dynamic-static hedging}

\author{Igor Halperin\thanks{Quantitative Research, JPMorgan Chase,
277 Park Avenue, New York, NY 10172, USA, email: igor.halperin@jpmorgan.com}, 
Andrey Itkin\thanks{Department of Finance and Risk Engineering, NYU Polytechnic Institute,
6 Metro Tech Center, RH 517E, Brooklyn NY 11201, USA, email: aitkin@poly.edy}\\ \\
\centerline{\small (Submitted to Quantitative Finance) }
}

\maketitle

\begin{abstract}
This work addresses the problem of optimal pricing and hedging of a European option on an illiquid asset Z using two proxies: a liquid asset S and a liquid European option on another liquid asset Y. We assume that the S-hedge is dynamic while the Y-hedge is static. Using the indifference pricing approach we derive a HJB equation for the value function, and solve it analytically (in quadratures) using an asymptotic expansion around the limit of the perfect correlation between assets Y and Z. While in this paper we apply our framework to an incomplete market version of the credit-equity Merton's model, the same approach can be used for other asset classes (equity, commodity, FX, etc.), e.g. for pricing and hedging options with illiquid strikes or illiquid exotic options.
\end{abstract}

\section{Introduction}
Consider a trader who wants to buy or sell a European option $ C_Z $ on asset $Z$
with maturity $ T $ and payoff $ G_Z $. The trader wants to hedge this position, but the underlying asset $Z$ is illiquid. However, some liquid proxies of $Z$ are available in the marketplace. First, there is a financial index (or simply an {\it index})  $S$ (such
as e.g. S\&P500 or CDX.NA)\footnote{Here we refer to this instrument as an index, but it could be any "linear" instrument such as stock, forward, etc.}  whose market price is correlated
with  $Z$. In addition, there is another correlated asset $Y$ which has a liquidly traded option $C_Y$ with a payoff $G_Y$ similar to that of $C_Z$, and with the same maturity $T$. The market price $p_Y$ of $C_Y$ is also known.

Our trader realizes that hedging $Z$-derivative with the index $S$ alone may not be sufficient for a number of reasons. First, she might be faced with a situation where correlation coefficients $\rho_{yz}, \rho_{sz}$ (which for simplicity are assumed to be constant) are such that $\rho_{yz} > \rho_{sz}$. In this case we would intuitively expect a better hedge produced by using $Y$ or $C_Y$ as the hedging instruments. Second, if we bear in mind a stochastic volatility-type dynamics for $Z$, the stochastic volatility process may be "unspanned", i.e. the volatility risk of the option may not be traded away by hedging in option's underlying\footnote{For a discussion of such scenarios for commodities markets, see \cite{TrolleSchwartz2009}.}. If that is the case, one might want to hedge the unspanned stochastic volatility by trading in a "similar" option with on the proxy asset $Y$. So our trader is contemplating a hedging strategy that would use both $ S $ and $ Y $. To capture an "unspanned" stochastic volatility, the trader wants to use a derivative $ C_Y $ written on $ Y $ rather than asset $ Y $ directly.

As transaction costs are usually substantially higher for options than for underlyings,
our trader sets up a static hedge in $ C_Y$ and a dynamic
hedge in $ S_t $. The static hedging strategy amounts to selling $ \alpha $ units of $ C_Y$
options at time $ t = 0 $. An {\it optimal} hedging strategy would be composed of
a pair $ (\alpha^*, \pi_s^*) $ where $ \alpha^* $ is the optimal static hedge, and
$ \pi_s^* $ (where $ 0 \leq s \leq T$) is an optimal dynamic hedging strategy in index $ S_t $. The pair $ (\alpha^*, \pi_s^*) $ should be obtained using a proper model. The same model
should produce the highest/lowest price for which the trader should agree to buy or sell the
$Z$-option.

In this paper we develop a model that formalizes the above scenario by supplementing it with the specific dynamics for asset prices $S_t, Y_t $ and $ Z_t $, and providing criteria of optimality for pricing options $C_Z$. For the former, we use a standard correlated log-normal dynamics. For the latter, we employ the utility indifference framework with an exponential utility, pioneered by
\citet{HN89,Davis97} and others, see e.g. \cite{HH2008} for a review.
As will be shown below, this results in a tractable formulations with analytical (in quadratures) expressions for optimal hedges and option prices.

As the above setting of pricing and hedging an illiquid option position using a pair of liquid proxies
(e.g. a stock and an option on a different underlying)
 is quite general, one could visualize its potential applications for various asset classes such as equities, commodities, FX etc.
For definiteness, in this paper we concentrate on a problem of practical interest for counterparty credit risk management \footnote{Most of the formulae below, excluding those that use specific forms of payoffs, are general and applicable for other similar settings.}.  Namely, we consider the problem of pricing and hedging an exposure to a counterparty with an illiquid debt, and in the absence of liquidly traded CDS referencing this counterparty. For such situation, no market-implied spreads are
available for the counterparty in question. Instead, one should rely on a model to come up with
{\it theoretical} credit spreads for the counterparty. To this end,
we use a version of the classical Merton equity-credit model \cite{m74} which is set up
in a multi-name setting,
and under the physical (i.e. "real", not "risk-neutral") measure. Most importantly, unlike the classical Merton's model, we do not intend to use firm's equity to hedge firm's debt. Instead, illiquid debt is hedged with a proxy liquid debt, and a proxy credit index.
In what follows, to differentiate our framework from that of the classical Merton model, we will
refer to it as the Hedged Incomplete-market Merton's Dynamics, or HIMD for short.

\subsection{Relation to previous literature}
Our model unifies three strands in the literature on indifference pricing.

The fist strand deals with hedging an option with a proxy asset, as
developed in \citet{Davis99, HH2002, MusielaZariphopoulou2004}, and others.
In this setting, one typically hedges an option on an illiquid underlying
with a liquid proxy asset.

The second strand develops generalizations of the classical
Merton credit-equity model to an incomplete market
setting. Typically, this is achieved by de-correlating asset value and
equity price at the level of a single
firm, see e.g. \citet{Leung2008, Jaim2009}. As long as we do not use
firms equity to hedge firm's debt
but instead use a liquid proxy bond as a hedge, such modification of
the Merton model is not needed in our setting.

The third strand is presented by \cite{IlhanSircar2006} who develop
a static-dynamic indifference hedging approach for barrier options. Ilhan
and Sircar considers hedging a barrier option under stochastic
volatility using static hedges in vanilla options on the same underlying
plus a dynamic hedge in the underlying. This results in a two-dimensional
Hamilton-Jacobi-Bellmann (HJB) equation. Our construction is similar but
our hedges are a proxy asset and a proxy option, while volatility is
taken constant for simplicity.

\section{Static hedging in indifference pricing framework}
Borrowing from an approach of \cite{IlhanSircar2006} for a similar (but not identical) setting, we now show how the method of indifference utility pricing can be generalized to incorporate our scenario of a mixed dynamic-static hedge.

To this end, let $\Pi(Y_T, Z_T)$ be the final payoff of the portfolio consisting of our option positions, i.e.
\begin{equation} \label{Payoff}
\Pi^{\alpha}(Y_T, Z_T) = G_Z - \alpha^* G_Y
\end{equation}
As long as both European options $C_Z, C_Y$ pay at the same maturity $T$, we can view this as the payoff of a combined ("static hedge portfolio") option $g_Z^{\alpha}$, which involves payoffs
$ G_Z $ and $ G_Y$ of both derivatives $C_Z$ and $C_Y$. Such option may be priced using the standard utility indifference principle. The latter states that the derivative price $ g_Z^{\alpha}$ is such that the investor should be indifferent to the choice between two investment strategies. With the first strategy, the investor adds the derivatives to her portfolio of bonds and stocks
(or indices\footnote{The stock is equivalent to our index $S$ in the setting of the Merton's optimal investment problem.}) $S$, thus taking $g_Z^{\alpha}$ from, and adding $\alpha p_Y $ to her initial cash $x$. With the second strategy, the investor stays with the optimal portfolio containing bonds and the stocks/indices.

The value of each investment is measured in terms of the {\it value function} defined as the conditional expectation of utility $U(W_T)$ of the terminal wealth $W_T$ optimized over trading strategies. In this work, we use an exponential utility function
\begin{equation} \label{expUtility}
U(W) = - e^{ - \gamma W}
\end{equation}
\noindent where $\gamma$ is a risk-aversion parameter. In our case, the terminal wealth is given by
the following expression:
\begin{equation*}
W_T = X_T + \Pi^{\alpha} (Y_T, Z_T)
\end{equation*}
\noindent with $X_T$ be the total wealth at time $T$ in bonds and index $ S $. In turn, the value function reads
\begin{equation} \label{value_function}
V(t,x,y,z) = \sup_{ \pi_t \in \mathcal{M} } \mathbb{E} \left[ U \left(
 X_T + \Pi^{\alpha} (Y_T, Z_T) \right) \Big| X_t = x, Y_t = y, Z_t = z \right]
\end{equation}
\noindent where $ \mathcal{M} $ is a set of admissible trading strategies that require holding of initial cash $x$. The expectation in the \eqref{value_function} is taken under the ``real-world''
measure $\mathds{P}$.

For a portfolio made exclusively of stocks/indices and bonds, the value function for the exponential utility is known from the classical Merton's work:
\begin{equation} \label{Merton_vv}
V^0(x,t) = - e^{ - \gamma x e^{r \tau} - \frac{1}{2} \eta_s^2 \tau}
\end{equation}
\noindent where $\tau = T - t$, $r$ is the risk free interest rate assumed to be constant, and $\eta_s =
(\mu_s - r)/\sigma_s $ is the stock Sharpe ratio.

In our setting, in addition to bonds and stocks/indices, we want to long $C_Z$ option and short $\alpha$ units of $C_Y$ option to statically hedge our $C_Z$ position, or, equivalently, buy the $g_z^{\alpha}$ option.

The value function in our problem of optimal investment in bonds, index and the composite option $g_z^{\alpha} $ has the following form:
\begin{equation} \label{valueFunction}
V(x,y,z,t) = \sup_{\pi_t \in \mathcal{M}} E\left( - e^{-\gamma
( X_T + \Pi^{\alpha}(Y_T, Z_T)} \Big| X_t = x, Y_t = y, Z_t = z\right)
\end{equation}
\noindent where $X_T$ is a cash equivalent of the total wealth in bonds and the index at time $T$.
We represent it in a form similar to \eqref{Merton_vv}:
\begin{equation} \label{vv_HIM}
V(x,y,z,t) = - e^{ - \gamma x e^{r \tau}- \frac{1}{2} \eta_s^2 \tau} \Phi(y,z,\tau)
\end{equation}
\noindent where function $\Phi$ will be calculated in the next sections. The indifference pricing equation reads
\begin{equation*}
V(x,y,z,t) = V^0(x + g_Z^{\alpha} - \alpha p_Y,t)
\end{equation*}
Plugging this in \eqref{Merton_vv} and \eqref{vv_HIM} and re-arranging terms, we obtain
\begin{equation*}
g_Z^{\alpha} = - \frac{1}{\gamma} e^{ r \tau} \log \Phi(y,z,\tau) + \alpha p_Y
\end{equation*}
The highest price of the $Z$-derivative is given by choosing the optimal static hedge given by the number $\alpha$ of the $Y$-derivatives, i.e.
\begin{align} \label{indif_price_2}
g_Z^{\alpha^*} &= - \frac{1}{\gamma} e^{r \tau} \log \Phi_{\alpha^*}(y,z,\tau) + \alpha^{*} p_Y \\
\alpha^* &= \arg \max_{\alpha} \left\{ - \frac{1}{\gamma} e^{r \tau} \log \Phi_{\alpha}(y,z,\tau)
+ \alpha p_Y \right\} \nonumber
\end{align}
\noindent where we temporarily introduced subscript $\alpha$ in $\Phi_{\alpha}$ to emphasize that the value function depends on $\alpha$ through a terminal condition.

\section{The HJB equation for HIMD}
To use \eqref{indif_price_2} and thus be able to compute both the option price and optimal static hedge, we need to find the "reduced" value function $\Phi$. To this end, we first derive the Hamilton-Jacobi-Bellman (HJB) equation for our model, and then obtain its analytical (asymptotic) solution.

Let $\pi = \pi_t(x)$ be the dynamic investment
strategy in the index $ S_t $
at time $t$ starting with the initial cash $x$, and
$\mathcal{L}^\pi$ be the Markov generator of price dynamics corresponding
to strategy  $\pi $.
Both the optimal dynamic
strategy and the value function
should be obtained as a solution of the HJB equation
\begin{equation} \label{HJB}
V_t + \sup_{\pi} \mathcal{L}^\pi V = 0
\end{equation}

We assume that all state variables $S_t, Y_t, Z_t$ follow a
geometric Brownian motion process with constant drifts $\mu_i$ and
volatilities $\sigma_i, i \in (x,y,z)$
\begin{align*}
d S_t = \mu_x S_t dt + \sigma_x S_t d W_t^{(x)} \\
d Y_t = \mu_y Y_t dt + \sigma_y Y_t d W_t^{(y)} \\
d Z_t = \mu_x Z_t dt + \sigma_z Z_t d W_t^{(z)} \\
\end{align*}
If our total wealth at time $ t $ is $ X_t = x $ and we invest amount $ \pi $
of this wealth into the index and the rest in a risk-free bond, the stochastic differential equation
for $ X_t $ is obtained as follows:
\begin{equation*}
dX_t = r \left( X_t - \pi \right) dt + \frac{\pi}{S_t} dS_t = \left( r X_t + \pi \sigma_x \eta_s \right)
dt + \pi \sigma_s dW_t^{(x)} \; , \quad \eta_s = \frac{\mu_x -r}{\sigma_x}
\end{equation*}
Then $\mathcal{L}^\pi$ reads
\begin{align*}
\mathcal{L}^\pi &=  \left( r x + \pi (\mu_x - r) \right) V_x
+ \frac{1}{2}\sigma^2_s \pi^2 V_{xx} + \mu_y y V_y + \frac{1}{2}\sigma^2_y y^2 V_{yy} +
+ \mu_z z V_z \\
&+ \frac{1}{2}\sigma^2_z z^2 V_{zz} + \rho_{xy} \sigma_x \sigma_y \pi y V_{xy} + \rho_{xz} \sigma_x \sigma_z \pi z V_{xz}
+ \rho_{yz} \sigma_y \sigma_z y z V_{yz}, \nonumber
\end{align*}
\noindent where $V(x,y,z,t)$ is defined on the domain $\mathbb{R}(x,y,z,t): [0, \infty) \times [0, \infty) \times [0, \infty) \times [0,T]$.

Since $\mathcal{L}^\pi$ is a regular function of $\pi$, $\sup_{\pi}$ is achieved at
\begin{equation*}
\pi_s^*(x) = - \frac{ \eta_s V_x + \rho_{xy} \sigma_y y V_{xy} + \rho_{xz} \sigma_z z V_{xz}}{\sigma_x V_{xx}}
\end{equation*}
Plugging this into \eqref{HJB}, we obtain
\begin{align} \label{PDE}
V_t &+ r x V_x + \mu_y y V_y + \frac{1}{2}\sigma^2_y y^2 V_{yy} + \mu_z z V_z + \frac{1}{2}\sigma^2_z z^2 V_{zz} +
\rho_{yz} \sigma_y \sigma_z y z V_{yz} \\
&- \frac{1}{2} \frac{ (\eta_s V_x + \rho_{xy} \sigma_y y V_{xy} + \rho_{xz} \sigma_z z V_{xz})^2}{V_{xx}} = 0 \nonumber
\end{align}
This is a nonlinear PDE with respect to the dependent variable $V(t,x,y,z)$ with standard boundary conditions (see \cite{MusielaZariphopoulou2004})
and the terminal condition determined by a choice of the writer's
maximal expected utility (value function) of the terminal wealth $W_T$.

Note that so far the derivation is valid for a generic utility function.
To make further progress we specialize to the case of exponential
utility in \eqref{expUtility} since it gives rise to a natural dimension
reduction of the HJB equation. Indeed, the ansatz
\begin{equation} \label{dimRed}
V(t,x,y,z) = - \exp\left(-\gamma x e^{r(T-\tau)}\right) G(h,s,\tau)
\end{equation}
\noindent with $h = \log (y/K_y), s = \log (z/K_z)$ is both consistent with terminal condition \eqref{valueFunction}
and, upon substitution in (\ref{PDE}), leads to a PDE for function $ G $ which does not
contain variable $ x $:
\begin{align} \label{GpdeWS}
    G_{\tau} &= \hat{\mu}_y G_h + \hat{\mu}_z G_s + \frac{1}{2}\sigma^2_y G_{hh} + \frac{1}{2}\sigma^2_z G_{ss} + \rho_{yz} \sigma_y \sigma_z G_{hs} \\
& - \frac{1}{2} \eta_s^2 G - \frac{1}{2}
\frac{ ( \rho_{xy} \sigma_y G_{h} + \rho_{xz} \sigma_z G_{s})^2}{G}. \nonumber
\end{align}
Here
\begin{equation*}
\hat{\mu}_y = \mu_y -\frac{1}{2}\sigma^2_y - \eta_s \rho_{xy} \sigma_y \, , \;
\hat{\mu}_z = \mu_z -\frac{1}{2}\sigma^2_z - \eta_s \rho_{xz} \sigma_z
\end{equation*}
Equation \eqref{GpdeWS} is defined at the domain $\mathbb{R}(h,s,t): [-\infty, \infty) \times [-\infty, \infty) \times [0,T]$. The initial condition for this equation is obtained from \eqref{valueFunction}.

In what follows, we choose a specific payoff of the form \eqref{Payoff} with $\Pi_Y = \min(Y,K_y), \ \Pi_Z = \min(Z,K_z)$ that corresponds to a portfolio of bonds of firms $ Y $ and $ Z $ with notionals $K_y, K_z$ within the Merton credit-equity model. Then the terminal condition for $G(h,s,\tau)$ reads
\begin{equation} \label{initCondWS}
G(h,s,0) = \exp \left[ - \gamma \left( K_z e^{s_{-}} - \alpha K_y e^{h_{-}} \right) \right]
\end{equation}
\noindent where $ s_{-} = \min(s,0) $ and $ h_{-} = \min(h,0) $.

\section{Asymptotic solutions of \eqref{GpdeWS}}
We were not able to find a closed form solution of \eqref{GpdeWS} with the initial condition  \eqref{initCondWS}. On the other hand, a numerical solution of this equation is expensive, especially when it should be used many times for calibration to market data. Therefore, we proceed with aasymptotic solutions of \eqref{GpdeWS}. We suggest two approaches to construct asymptotic solutions.

\subsection{First method}
As we want to statically hedge option $C_Z$ with options on another underlying, we look for an asset $ Y $ that is strongly correlated with asset $ Z $. Further,  if we have a ``similar'' option $ C_Y $ on asset $ Y $ (i.e. similar maturity,type, strike, etc.), we expect that such option provides a good static hedge for our option $C_Z$ \footnote{As an example, we mention the case of equity options referencing the same underlying, i.e. $Y=Z$, but $K_z \ne K_y$. We may want to hedge an illiquid option with strike $K_z$ (say, deep OTM) with a liquid option on the same underlying but with a different strike $K_y$. Under this setup, we have $\rho_{yz} = 1$, i.e. a prefect correlation case.}.

Therefore, a natural assumption would be to consider $1 - \rho_{yz}$ to be a small parameter under our setup. Utilizing this idea we represent the solution of \eqref{GpdeWS} as a formal perturbative expansion in powers of $ \varepsilon $:
\begin{equation} \label{expan}
G = \sum_{i=0}^\infty \varepsilon^i G_i,
\end{equation}
\noindent where $\varepsilon$ is the small parameter to be precisely defined in the next section.

As we shown below, \eqref{GpdeWS} can be solved analytically (in quadratures) to any order of this expansion, thus significantly reducing the computation time.

\subsubsection{The HJB equation in "adiabatic" variables}
We start with a change of variables $(h,s) \rightarrow (w,v) $ defined as follows:
\begin{align} \label{uv}
w &= h\frac{1}{\sigma_y} + \tau \frac{\hat{\mu}_y}{\sigma_y} \\
v &= - h\frac{1}{\sigma_y} + s\frac{\rho_{xy}}{\rho_{xz} \sigma_z}
+ \tau \left( -\frac{\hat{\mu}_y}{\sigma_y}
+ \frac{\hat{\mu}_z}{\sigma_z} \frac{\rho_{xy}}{\rho_{xz}}
\right) \nonumber
\end{align}
Simultaneously, we change the dependent variable $ G \rightarrow \Phi $ as follows:
\begin{equation} \label{changeOfFun1}
G(h,s, \tau) = e^{ - \frac{1}{2} \eta_s^2 \tau} \Phi(w,v,\tau)
\end{equation}
Using \eqref{GpdeWS}, \eqref{uv} and \eqref{changeOfFun1}, we obtain the following PDE for function $ \Phi $:
\begin{equation} \label{Phi_uv}
\Phi_{\tau} = \frac{1}{2} \Phi_{ww} + \frac{1}{2} \frac{\rho_{xz}^2 - 2 \rho_{xy} \rho_{xz} \rho_{yz}
+ \rho_{xy}^2}{\rho_{xz}^2} \Phi_{vv} + \frac{ \rho_{xz} - \rho_{xy} \rho_{yz}}{ \rho_{xz}}
\Phi_{wv} - \frac{1}{2} \rho_{xy}^2 \frac{ \Phi_w^2}{\Phi}.
\end{equation}
Further we will show that $v$ is a slow ("adiabatic") \footnote{See e.g. \cite{HairerLubichWanner2006}.} variable of our asymptotic method, while $w$ becomes a "fast" variable.

In what follows we need an inverse of \eqref{uv} at $\tau=0$:
\begin{equation*}
h = \sigma_y w \; , \; \;
s = (v+w)\frac{\rho_{xz}\sigma_z}{\rho_{xy}}
\end{equation*}
Using this in \eqref{initCondWS}, we obtain the initial condition in $(w,v) $ variables for the function $\Phi(w,v,\tau)$:
\begin{equation} \label{initCondUV}
\Phi(w,v,0) = \exp \left[ - \gamma \left(
K_z e^{\frac{\rho_{xz}}{\rho_{xy}} \sigma_z \left( v + w \right)_{-}} -
\alpha K_y e^{ \sigma_y w_{-}} \right) \right]
\end{equation}
where $ (x)_{-} = \min(x, 0) $ for any real $ x $.
\subsubsection{Cosine law in 3D and "adiabatic" limit}
Recall that a correlation matrix $ \Sigma $ of $ N $ assets
can be represented as a Gram matrix with matrix elements $ \Sigma_{ij} = \langle {\bf x}_i , {\bf x}_j \rangle $ where $ {\bf x}_i, \, {\bf x}_j $ are
unit vectors on a $ N-1 $ dimensional hyper-sphere $ S^{N-1} $. Using
the 3D geometry, it is easy to establish the following {\it cosine law}
for correlations between three assets:
\begin{equation}  \label{cosine}
\rho_{xy} = \rho_{yz} \rho_{xz} + \sqrt{ \left(1 - \rho_{yz}^2 \right) \left(1 - \rho_{xz}^2 \right) }
\cos ( \phi_{xy}),
\end{equation}
\noindent with $\phi_{xy}$ being an angle between
$ {\bf x} $ and its projection on the plane spanned by $ {\bf y}, {\bf z} $.

As discussed e.g. by \cite{Dash2004}, three variables $\rho_{xy}, \rho_{yz}, \phi_{xz}$ are independent, but $ \rho_{xy}, \rho_{yz}, \rho_{xz} $
are not. Therefore, one of them, e.g. $\rho_{xy}$, has to be
found using \eqref{cosine} given $\rho_{xz}, \rho_{yz}, \phi_{xy}$.

Further we define $\varepsilon$ as
\begin{equation} \label{eps}
\varepsilon = \sqrt{1 - \rho_{yz}^2} \ll 1,
\end{equation}
\noindent and also define the following constants
\begin{equation} \label{theta_12}
\theta_1 =  \frac{\sqrt{ 1 - \rho_{xz}^2}}{\rho_{xz}} \cos( \phi_{xy} ),
\qquad \theta_2 = 1 + \theta_1^2
\end{equation}
Using \eqref{cosine} and definitions in \eqref{eps}, \eqref{theta_12}, coefficients at $\Phi_{uv} $ and $\Phi_{vv}$ in \eqref{Phi_uv} are evaluated as follows:
\begin{align*}
-1 &+ \frac{\rho_{xy} \rho_{yz} }{\rho_{xz}} = \varepsilon \theta_3, \quad
\rho_{xy} = \rho_{xz}\beta, \quad
\frac{\rho_{xz}^2 - 2 \rho_{xy} \rho_{xz} \rho_{yz} + \rho_{xy}^2}{\rho_{xy}^2} = \varepsilon^2\theta_2 \\
\beta &= \sqrt{1 - \varepsilon^2} + \varepsilon \theta_1, \quad
\theta_3 = \sqrt{1-\varepsilon ^2} \theta_1 -\varepsilon.
\end{align*}
Accordingly using this notation \eqref{Phi_uv} takes the form
\begin{equation} \label{Phi_uv_2}
\Phi_{\tau} = \frac{1}{2} \Phi_{ww} + \varepsilon \theta_3 \Phi_{wv} + \frac{1}{2} \varepsilon^2 \theta_2 \Phi_{vv} - \frac{1}{2} \rho_{xz}^2 \beta^2 \frac{\Phi_w^2}{\Phi}
\end{equation}
In the limit $\varepsilon \rightarrow 0$ this equation
does not contain any derivatives wrt $v$, therefore $v$ enters the
equation only as a parameter (since $G(w,v,\tau)$ is a function of $v$). We
call this limit the {\it adiabatic limit} in a sense that will be
explained below.

It should be noted that our expansion in powers of $ \varepsilon $ can diverge
if $ \rho_{xy} $ is very small. We exclude such situations on the "financial" grounds assuming that all pair-wise correlations in the triplet $(S_t, Y_t, Z_t)$ are reasonably high (of the order of 0.4 or higher in practice), for our hedging set-up to make sense in the first place. Thus, parameter $ \theta_1$ is treated as $O(1)$\footnote{While parameter $\theta_2$ is always $O(1)$ and positive, parameter $\theta_1$ could be both positive and negative for typical values of correlations. For example, if $(\rho_{xy}, \rho_{xz}, \rho_{yz}) = (0.4, 0.4, 0.8) $, then $\theta_1 = 0.33$, while for $ (\rho_{xy}, \rho_{xz}, \rho_{yz}) = (0.3, 0.2, 0.8)$ it is $\theta_1 = -0.22$. The cosine law can also be used to find proper values of correlation parameters in the limit $ \rho_{yz} \rightarrow 1 $. To this end, we first use \eqref{cosine} to convert the estimated triplet $ (\rho_{xy}, \rho_{xz},\rho_{yz}) $ into a triplet of independent variables $ ( \rho_{xz}, \phi_{xy},\rho_{yz}) $, and then take the limit $ \rho_{yz} \rightarrow 1 $ while keeping $ \rho_{xz} $ and $ \phi_{xy} $ constant.}.

\subsection{Second approach}
It turns out that the last equation of the previous section could be further simplified. Introducing new independent variables
\begin{equation} \label{Uchange}
u = \frac{1}{\sqrt{\theta_2} \beta} \left(\theta_2 w - \frac{\theta_3}{\varepsilon} v \right),
\qquad \bar{v}= v/\varepsilon
\end{equation}
\noindent we can transform \eqref{Phi_uv_2} into the following equation
 \begin{equation} \label{Phi_vv_2}
\Phi_{\tau} = \frac{1}{2} \Phi_{uu} + \frac{1}{2} \theta_2 \Phi_{\bar{v}\bar{v}} - \frac{1}{2} \rho_{xz}^2 \theta_2 \frac{\Phi_u^2}{\Phi}
\end{equation}

It is seen that in new variables the mixed derivative drops from from the equation, as so does
$\varepsilon$. However,
further let us formally introduce a multiplier $\mu$ in the term $\Phi_{\bar{v}\bar{v}}$ which transforms  \eqref{Phi_vv_2} into
 \begin{equation} \label{Phi_vv_2mu}
\Phi_{\tau} = \frac{1}{2} \Phi_{uu} + \frac{1}{2}\mu \theta_2 \Phi_{\bar{v}\bar{v}} - \frac{1}{2} \rho_{xz}^2 \theta_2 \frac{\Phi_u^2}{\Phi}
\end{equation}
Let us also formally assume that $\mu$ is small under certain conditions. The idea of this trick is as follows. One way to construct an asymptotic solution of the \eqref{Phi_vv_2} is to assume that all derivatives are of the order $O(1)$, and then estimate all the coefficients. If one manages to  find a coefficient which is $O(\mu)$, then it is possible to build an asymptotic expansion using that coefficient (or $\mu$ itself) as a small parameter. If, however, all the coefficients in the considered PDE are of order $O(1)$ we need to check if perhaps some of the derivatives in the \eqref{Phi_vv_2} are small, e.g. $O(\mu)$. If this is the case, in order to apply standard asymptotic methods we formally have to add a small parameter $\mu$ as a multiplier to the derivative which is $O(\mu)$
\footnote{In other words write $\Phi_{\bar{v}\bar{v}}  = \mu (\Phi_{\bar{v}\bar{v}}/\mu) = \mu (\bar{\Phi}_{\bar{v}\bar{v}})$, where $\bar{\Phi}_{\bar{v}\bar{v}} \propto O(1)$}, make an asymptotic expansion on $\mu$, solve the obtained equations in every order on $\mu$, and at the end in the final solution put $\mu = 1$. That is exactly the way we want to proceed with.

This means that instead of the \eqref{expan}, we now have the following expansion:
\begin{equation} \label{expan1}
G = \sum_{i=0}^\infty \mu^i G_i.
\end{equation}

To find conditions when $\Phi_{\bar{v}\bar{v}}$ could be small as compared with the other terms in the \eqref{Phi_vv_2}, we use an inverse map at $\tau=0: \ (h,s) \rightarrow (u,\bar{v})$
\footnote{At $\varepsilon \rightarrow 0$ this is a regular map.}
\begin{align*}
s &= \frac{\sigma_z}{\sqrt{\theta _2}}\left(\bar{v} \frac{\theta_1}{\sqrt{\theta _2}} + u \right), \\
h &= \frac{\sigma_y}{\sqrt{\theta _2}}\left(\bar{v} \frac{\theta_3}{\sqrt{\theta _2}} + u \beta\right)
\nonumber
\end{align*}
\noindent and rewrite the payoff function \eqref{initCondWS} in the form
\begin{align} \label{initCondUv1}
\Phi(u,\bar{v},0) &= \exp \left[ - \gamma \left(K_z e^{\zeta \sigma _z (\omega_1 + u)_{-}}
- \alpha K_y e^{\zeta \sigma_y \beta (\omega_2 + u)_{-}} \right) \right], \\
\omega_1 &= \bar{v}\frac{\theta_1}{\sqrt{\theta_2}},
\qquad \omega_2 = \bar{v} \frac{\theta_3}{\beta \sqrt{\theta _2}},
\qquad \zeta = \frac{1}{\sqrt{\theta_2}} \nonumber
\end{align}

Suppose that $v \ge 0, \ u < -\omega_1$ or $v <0, \ u < -\omega_2$. Differentiating the payoff twice by $u$ and twice by $v$ and computing the ratio of the first and second terms in the rhs of the \eqref{Phi_vv_2}, one can see that in the limit $\varepsilon \rightarrow 0$ this ratio becomes $\mu = \theta_2 \Phi_{\bar{v}\bar{v}}/\Phi_{uu} = \theta_1^2$. Typical values $\rho_{xz} = 0.3, \rho_{xy} = 0.2, \rho_{yz} = 0.8$ give rise to $\mu = 0.05$, therefore the second term is small as compared with the first one, and $\mu$ is a good small parameter. This, however changes if $\rho_{xz}$ is small or/and $cos(\phi_{xy})$ is close to 1, and then $\mu \propto O(1)$. Still in this case we have $\varepsilon < 1$ which can be used as a small parameter. Therefore, our approach is as follows:
\begin{enumerate}
\item If $ \theta_1^2 \ll 1$ we use \eqref{Phi_vv_2} and find its asymptotic solutions using  \eqref{expan1}. This is better than using \eqref{Phi_uv_2} because first, $\mu$ is typically smaller then $\varepsilon$, and second, the term $\Phi_{\bar{v}\bar{v}}$ in our first method \eqref{expan} appears only in the second order of approximation while in the second method it is taken into account already in the first order on $\mu$.

\item If, however, $ \theta_1^2 \propto O(1)$, then $\mu$ is not anymore a small parameter, therefore we use \eqref{Phi_uv_2} and solve it asymptotically using \eqref{expan}.
\end{enumerate}

In general,  this argument cannot be applied if $\bar{v} \ge 0, \ u \ge -\omega_2$ or $v <0, \ u \ge -\omega_1$ because then both derivatives of the payoff vanish. However, the above argument
is intended to provide an intuition as to why $\Phi_{\bar{v}\bar{v}}$ could be much smaller that the other terms in the rhs of \eqref{Phi_vv_2}. This intuition can be verified numerically, and our test examples clearly demonstrate that smallness of $\mu$ often takes place. Below we discuss under which conditions this could occur.

Note that at the first glance, the described method looks similar to the quasi-classical approximation in quantum mechanics (\cite{LL3}). The similarity comes from the observation that transformation \eqref{Uchange} is singular in $\varepsilon$ which is similar to the quasi-classical limit $\hbar \rightarrow 0$. If we would construct an asymptotic expansion on $\varepsilon$ we would expand the rhs of the \eqref{Phi_vv_2} on $\varepsilon$, but not the payoff function. After getting the solution of \eqref{Phi_vv_2} in zero-order approximation on $\varepsilon$ as a function of $(u,\bar{v})$, we
would apply the inverse transformation $(u,\bar{v}) \rightarrow (h,s)$ which is non-singular. Therefore, the final result would not contain any singularity. Since $\mu = \theta_1^2$ is defined via $\rho_{xz}, \rho_{xy}$ and $\rho_{yz} = \sqrt{1-\varepsilon^2}$, it could seem that $\mu = \mu(\varepsilon)$, and we face a "quasi-classical" situation.

However, as explained above, the independent parameters are $\rho_{yz}, \rho_{xz}$ and $\cos(\phi_{xy})$. Therefore, by definition $\mu = \theta^2_1(\rho_{xz}, \cos(\phi_{xy}))$ doesn't depend on $\rho_{yz}$, or on $\varepsilon$. Thus, our two methods actually correspond to different assumptions. The first one utilizes a strong correlation between assets Z and Y. The second assumes a strong correlation between index $S$ and asset $Z$ while at the same time the vector of correlation $\rho_{xz}$ in 3D space is not collinear to the vector of correlation $\rho_{xy}$. By financial sense this means (see \eqref{cosine}) the following.
\begin{enumerate}
\item Either $\rho_{xz}$ is about 1 and, therefore, $\rho_{xy} \approx \rho_{yz}$ . In other words, index $S$ strongly correlates to asset Z, so $S$ is almost $Z$, therefore correlation of Y and Z ($\rho_{yz}$) is close to correlation of Y and X ($\rho_{xy}$). That, in turn, means that asset Z can be dynamically hedged with $S$, and extra static hedge with Y doesn't bring much value. In contrast, under the former assumption static hedge plays an essential role.
\item Or $\cos(\phi_{xy}) \ll 1$ which means that $\rho_{xy} \approx \rho_{xz} \rho_{yz}$. For instance, $\rho_{xz} = 0.4, \ \rho_{yz} = 0.6, \ \rho_{xy} = 0.24$. This is an interesting case, since it differs from two previously considered assumptions on high value of either $\rho_{yz}$ or $\rho_{xz}$. Indeed, all correlations could be relatively moderate while providing a smallness of $\theta_1$.
\end{enumerate}

In what follows, we describe in detail the asymptotic solutions for zero and first order approximations in $\mu$, and outline a generalization of our approach to an arbitrary order in $\mu$. Asymptotic solutions in $\varepsilon$ are constructed in a very similar way and are given in Appendix~\ref{A}. Also to make our notation lighter in the next section we will us $v$ instead of $\bar{v}$ since that should not bring any confusion.

\subsection{Zero-order approximation}
\label{sect:zeroorder}
In the zero order approximation we set $\mu = 0$, so that \eqref{Phi_vv_2} does not
contain derivatives wrt $v$:
\begin{equation} \label{zeroEq}
\Phi_{0,\tau} = \frac{1}{2} \Phi_{0,uu} - \frac{1}{2} \bar{\rho}_{xz}^2 \frac{ \left( \Phi_{0,u} \right)^2} {\Phi_0},
\end{equation}
\noindent where $\bar{\rho}^2_{xz} = \theta_2 \rho^2_{xz}$. Therefore, dependence of the solution on $v$ is determined by the terminal condition. In other words, our system changes along variable $u$, but it remains static (i.e. of the order of $\mu^2$ slow) in variable $v$. Using analogy with physics, we call this limit the {\it adiabatic limit}.

The last equation can be solved by a change of dependent variable (closely related to
the Hopf-Cole transform, see e.g. \citet{HH2002, MusielaZariphopoulou2004}):
\begin{equation} \label{distortion}
\Phi_0(\tau,u,v) = \left[ \phi(\tau,u,v) \right]^{1/(1- \bar{\rho}_{xz}^2)}
\end{equation}
\noindent which reduces \eqref{zeroEq} to the heat equation
\begin{equation} \label{linPDE2}
\phi_{\tau} = \frac{1}{2} \phi_{uu},
\end{equation}
\noindent subject to the initial condition $\phi_{u,v,0} = \Phi(u,v,0)^{1-\bar{\rho}_{xz}^2}$. The latter can be obtained from the \eqref{initCondUv1} if one replaces $\gamma$ with the "correlation-adjusted" risk
aversion parameter $\bar{\gamma} = \gamma (1 - \bar{\rho}_{xz}^2)$. It can also be written as a piece-wise analytical function having a different form in different intervals of $u$-variable. If $ v \geq 0 $, we have
\begin{align} \label{init_cond_quand2}
\phi(u,v,0) =
\begin{cases}
\exp \left[ - \bar{\gamma} \left(K_z e^{\zeta \sigma_z \left(\omega_1 + u \right)} - \alpha K_y e^{ \zeta \beta \sigma_y\left(\omega_2 + u\right)} \right) \right],
& u < - \omega_1 \\
\exp\left[ - \bar{\gamma} \left(K_z - \alpha K_y e^{ \zeta \beta \sigma_y\left(\omega_2 + u\right)}\right) \right],
& - \omega_1 \le u < -\omega_2 \\
\exp\left[ - \bar{\gamma} \left(K_z - \alpha K_y\right) \right],
& u \ge -\omega_2,
\end{cases}
\end{align}
\noindent while for $v < 0$ we have
\begin{align*}
\phi(u,v,0) =
\begin{cases}
\exp \left[ - \bar{\gamma} \left(K_z e^{\zeta \sigma_z \left(\omega_1 + u \right)} - \alpha K_y e^{ \zeta \beta \sigma_y\left(\omega_2 + u\right)} \right) \right],
& u < -\omega_2 \\
\exp \left[ - \bar{\gamma} \left(K_z e^{\zeta \sigma_z\left(\omega_1 + u \right)} - \alpha K_y \right) \right],
& -\omega_2 \le u < - \omega_1 \\
\exp\left[ - \bar{\gamma} \left(K_z - \alpha K_y \right) \right],
& u \ge - \omega_1 \\
\end{cases}
\end{align*}

Using the well-known expression for the Green's function of our heat equation
$G_0 (u' - u, \tau) = \frac{ e^{-\frac{(u' - u)^2}{2 \tau}}}{\sqrt{2 \pi \tau}}$ (see e.g. \cite{Polyanin2002}),
the solution of \eqref{linPDE2} is then
\begin{equation*}
\phi(u,v,\tau) = \frac{1}{{\sqrt{2 \pi \tau}}} \int_{-\infty}^{\infty} e^{-\frac{(u' - u)^2}{2 \tau}} \phi(u',v,0) d u'
\end{equation*}
The explicit zero-order solution thus reads
\begin{equation} \label{Phi_0_sol}
\Phi_0(u,v,\tau) =
\left[\frac{1}{\sqrt{2 \pi \tau}}\int_{-\infty}^{\infty} e^{-\frac{(u' - u)^2}{2 \tau}} \phi(u',v,0) d u'\right]^{1/(1-\bar{\rho}_{xz}^2)}
\end{equation}
Note that \eqref{Phi_0_sol} provides the general zero-order solution for the HJB equation with arbitrary initial conditions at $ \tau = 0 $. For our specific initial conditions \eqref{init_cond_quand2}, the
solution is readily obtained in closed form in terms of the error (or normal cdf)  function (see Appendix \ref{A1}). However, for numerical efficiency it might be better to use another method
which is based on a simple observations that the expression in square brackets in the \eqref{Phi_0_sol} is just a Gauss transform of the payoff. This transform can be efficiently computed
using a Fast Gauss Transform algorithm which in our case is $O(2N)$ with $N$ being the number of grid points in $u$ space.

\subsection{First-order approximation}
\label{sect:Firstorder}
For the first correction in \eqref{Phi_vv_2} we obtain the following PDE
\begin{equation} \label{first2}
\Phi_{1,\tau} = \frac{1}{2} \Phi_{1,uu} - \bar{\rho}_{xz}^2 \frac{\Phi_{0,u}}{\Phi_0} \Phi_{1,u}
+ \frac{1}{2} \bar{\rho}_{xz}^2 \left( \frac{\Phi_{0,u}}{\Phi_0} \right)^2 \Phi_{1} + \Theta_1(u,v,\tau)
\end{equation}
\noindent with $\Theta_1(u,v,\tau) = \theta_2 \Phi_{0,vv}/2$.
This is an inhomogeneous {\it linear} PDE with variable coefficients. As long as our zero-order solution of \eqref{zeroEq} already satisfies the initial condition, this equation has to be solved subject to
the zero initial condition. This considerably simplifies the further construction.

We look for a solution to Eq.~(\ref{first2}) in the form
\begin{equation*}
\Phi_1(u,v,\tau) = \left[ \Phi_0(u,v,\tau) \right]^{\bar{\rho}_{xz}^2} H(u,v, \tau)
\end{equation*}
\noindent This gives rise to an inhomogeneous heat equation for function $H$ subject to zero initial condition
\begin{equation*}
H_{\tau} = \frac{1}{2} H_{uu} + \Theta_1\Phi_{0}^{-\bar{\rho}^2_{xz}}.
\end{equation*}
Thus, using the Duhamel's principle (\cite{Polyanin2002}) we obtain
\begin{equation} \label{FirstOrderSol}
\Phi_1(u,v,\tau) = \Phi_0^{\bar{\rho}_{xz}^2} (u,v,\tau)\int_0^{\tau} d\chi \int_{-\infty}^\infty
d u' \frac{e^{-\frac{(u- u' )^2}{2 (\tau-\chi)}} }{\sqrt{2 \pi (\tau-\chi)}} \Theta_1(u',v,\chi) \Phi_0^{-\bar{\rho}_{xz}^2}(u',v,\chi) \\
\end{equation}

There exists a closed form approximation of the internal integral (see Appendix \ref{B2}).

\subsection{Second order approximation and higher orders}
\label{sect:Secondorder}
The second order equation has the same form as the \eqref{first2}
\begin{equation*}
\Phi_{2,\tau} = \frac{1}{2} \Phi_{2,uu} - \bar{\rho}_{xz}^2 \frac{\Phi_{0,u}}{\Phi_0} \Phi_{2,u}
+ \frac{1}{2} \bar{\rho}_{xz}^2 \left( \frac{\Phi_{0,u}}{\Phi_0} \right)^2 \Phi_{2} + \Theta_2(u,v,\tau)
\end{equation*}
\noindent where
\begin{equation*}
\Theta_2(u,v, \tau) = \frac{1}{2}\theta_2 \Phi_{1,vv} - \frac{1}{2}\bar{\rho}_{xz}^2
\left[ \frac{\Phi_1^2 \Phi_{0,u}^2}{\Phi_0^3} - 2\frac{\Phi_1 \Phi_{0,u} \Phi_{1,u}}{\Phi_0^2}
+ \frac{\Phi_{1,u}^2}{\Phi_0} \right]
\end{equation*}
As this equation has to be solved also subject to zero initial conditions, the solution is obtained in the same way as above:
\begin{equation*}
\Phi_2(u,v,\tau) = \Phi_0^{\bar{\rho}_{xz}^2}(u',v,\tau) \int_0^{\tau}\int_{-\infty}^\infty  \frac{e^{-\frac{(u-u' )^2}{2 (\tau-\chi)}} }{\sqrt{2 \pi (\tau-\chi)}} \Theta_2(u,v,\chi) \Phi_0^{-\bar{\rho}_{xz}^2}(u',v,\chi) d\chi d u'
\end{equation*}
This shows that in higher order approximations in $\mu$ both
the type of the equation and boundary conditions stay the same. Therefore, the solution to the $n$-th order approximation reads
\begin{equation*}
\Phi_n(u,v,\tau) =  \Phi_0^{\bar{\rho}_{xz}^2}(u',v,\tau) \int_0^{\tau}\int_{-\infty}^\infty \frac{e^{-\frac{(u-u' )^2}{2 (\tau-\chi)}} }{\sqrt{2 \pi (\tau-\chi)}} \Theta_n(u,v,\chi) \Phi_0^{-\bar{\rho}_{xz}^2}(u',v,\chi) d\chi d u'
\end{equation*}
\noindent where $\Theta_n$ can be expressed via already found solutions of order $i, i=1...n-1$ and their derivatives on $u$ and $v$. The exact representation for $\Theta_n$ follows combinatorial rules and reads
\begin{equation} \label{Theta_n}
\Theta_n(u,v,\tau) = \frac{1}{2}\theta_2 \Phi_{n-1, vv} \nonumber - \frac{1}{2} \bar{\rho}_{xz}^2 \Xi_n,
\end{equation}
\noindent where $\Xi_n$ is a coefficient at $\mu^{n-1}, \ n > 1$ in the expansion of
\begin{align*}
& \frac{\Phi_{0,u}^2}{\Phi_0}\left[ \frac{(1 + \beta_1(\mu))^2}{1 +  \beta_2(\mu)} - \Phi_{n} +
2\frac{\Phi_{0}}{\Phi_{0,u}} \Phi_{n,u} \right], \\
\beta_1 &= \sum_{i=1}^\infty \mu^i \Phi_{i,u}/\Phi_{0,u}, \qquad
\beta_2 = \sum_{i=1}^\infty \mu^i \Phi_{i}/\Phi_{0}
\end{align*}
\noindent in series on $\mu$. This could be easily determined using any symbolic software, e.g. Mathematica. For instance $\Xi_3$ reads
\begin{equation*}
\Xi_3 = \frac{\left(-\Phi _1 \Phi _{0,u}+\Phi _0 \Phi _{1,u}\right) \left(\Phi _1^2 \Phi _{0,u}-\Phi _0 \Phi _1 \Phi _{1,u}+2 \Phi _0 \left(-\Phi _2 \Phi _{0,u}+\Phi _0 \Phi _{2,u}\right)\right)}{\Phi _0^4}
\end{equation*}
The explicit representation of the solutions of an arbitrary order in quadratures is important because, per our definition of $\mu$,
convergence of \eqref{expan} is expected to be relatively slow.
Indeed, if one wants the final precision to be about $O(0.1)$
at $\rho_{yz} = 0.8$ ($\mu \approx 0.36$), the number of important
terms $m$ in expansion \eqref{expan1}
could be rawly calculated as $\mu^{m+1} = 0.1$, which gives $m=1.25$, while
at precision $0.01$ this yields $m = 3.5$.

Note that all integrals with $n > 1$ do not admit a closed form
representation and have to be computed numerically. Again, this could be
done in an efficient manner using the Fast Gauss Transform.

\section{Validation of the method and some examples}

To verify quality of our asymptotic method we compare two sets of results. One is obtained using zero and first order approximations (being computed via a series representation and $\Omega$ functions given in Appendix~\ref{A1} in
\eqref{Jdef}, \eqref{Omega} and \eqref{Omega1}, or using the
Fast Gaussian Transform). Our tests showed that the number of terms in
the double sum that should be kept is small, namely truncating the
upper limit in $i$ from infinity to $i_{max} = 10$ produces nearly
identical results. Therefore, the total complexity of calculation is
about 45 computations of $\exp$ and $\mbox{Erfc}$ functions which
is very fast. A typical time required for this at a standard PC
with the CPU frequency 2.3 Ghz ranges from 0.68 sec (Test 1) to
0.35 sec (Test 2) (see below).

The other test is performed using a numerical solution of
\eqref{Phi_vv_2}. In doing so we use an implicit finite difference
scheme built in a spirit of \cite{KhiariOmrani2011_2}. After the
original non-linear equation is discretized to obtain the value
function at the next time level, we need to solve a 2D algebraic
system of equations each of which contains a non-linear term. This
could be done e.g. by applying a fixed point
iterative method (\cite{SallehZomayaBakar2007}). In other words, at
the first iteration as an initial guess we plug-in into the non-linear
term the solution obtained at the previous level of time. This reduces
the equation to a linear one since the non-linear term is explicitly
approximated at this iteration. Next we solve the resulting 2D system of
equations with a block-band matrix using a 2D LU factorization. At the
second iteration, the solution obtained in such a way is substituted
into the non-linear term again, so again it is approximated explicitly.
Then the new system of linear equations is solved and the new
approximation of the solution of the original
non-linear equation is obtained. We continue this
process until it converges. The number of iterations to needed for
numerical convergence
depends on gradients of the value function, which are considerably
influenced by the value of $\gamma$. For small values of
$\gamma$ (about 0.03) we need about 1-2 iterations, while for
$\gamma \approx 0.3$, 5-7 iterations might be necessary.
For higher values of $\gamma$, the fixed point iteration scheme
could even diverge, so another method has to be used instead. It is
also important to note that we solve the non-linear equation using the
dependent variable $\log(\Phi_0)$, rather than $\Phi_0$ to reduce relative
gradients of the solution.

Note that for {\it linear} 2D parabolic equations with mixed derivatives
more
efficient splitting schemes exist, see e.g. \cite{HoutWelfert2007}.
In principle, such schemes could be adopted to solve \eqref{Phi_vv_2}.
Though \eqref{Phi_vv_2} is a {\it non-linear} equation,
the presence of the non-linear term does not change its type (it is still
a parabolic equation), and second, does not affect
stability of the scheme if we approximate it implicitly and solve
the resulting non-linear algebraic equations. A more detailed
description of this modification of the splitting scheme
of Hout and Welfert will be presented elsewhere.

\begin{figure}
\begin{minipage}[b]{0.4\textwidth}
\fbox{\includegraphics[width=3in]{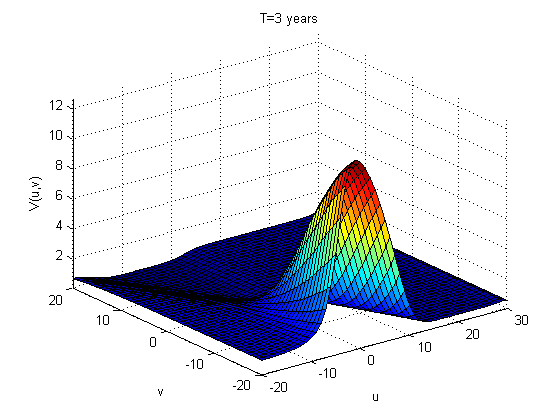}}
\end{minipage}
\hspace{0.1\textwidth}
\begin{minipage}[b]{0.4\textwidth}
\fbox{\includegraphics[width=3in]{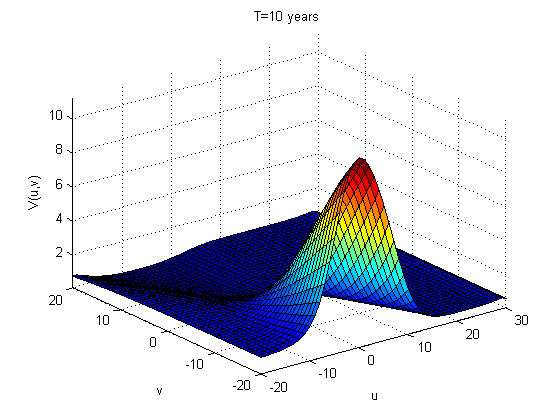}}
\end{minipage}
\caption{FD solution for the value function $V(u,v)$ at $T$=3 yr and $T$=10 yrs. The initial parameters are given in Table~\ref{Tab}, test 1.}
\label{TestUV3D}
\end{figure}

Implementation of the numerical algorithm is done similar to \cite{ItkinCarrBarrierR3}. In typical tests we use a non-uniform finite difference grid in $u$ and $v$ of size 50x50 nodes, where $-50 < u < 50, \ -30 < v < 30$. The number of steps in time depends on maturity $T$, because we use a fixed time step $\delta t =$ 0.1 yrs. Typical computational time for $T=3$ yrs on the same PC at $\gamma$ = 0.03 is 17 sec.
In Fig~\ref{TestUV3D} a 3D plot of the value function $V(u,v)$ is presented for the initial parameters marked in Table~\ref{Tab} as 'Test 1'. We also use $\gamma = 0.03, \ \alpha = 1$. It is seen that $V(u,v)$ quickly goes to constant outside of a narrow region around $u=0$ and $v=0$.
\begin{table}
\begin{center}
\begin{tabular}{|c|c|c|c|c|c|c|c|c|c|c|c|c|c|c|}
\hline
Test & $\mu_x$ & $\sigma_x$ & $r$ & $\rho_{yz}$ & $K_z$ & $\mu_z$ & $\sigma_z$ & $\rho_{xz}$ & $z_0$ & $K_y$ & $\mu_y$ & $\sigma_y$ & $\rho_{xy}$ & $y_0$  \cr
\hline
 1 & 0.04 & 0.25 & 0.02 & 0.8 & 110 & 0.05 & 0.2 & 0.4 & 100 & 90 & 0.03 & 0.3 & 0.3 & 100  \cr
 \hline
 2 & 0.04 & 0.25 & 0.02 & 0.8 & 110 & 0.05 & 0.3 & 0.3 & 50 & 90 & 0.03 & 0.3 & 0.2 & 100  \cr
 \hline
\end{tabular}
\end{center}
\caption{Initial parameters used in test calculations.}
\label{Tab}
\end{table}

\begin{figure}
\begin{minipage}[b]{0.4\textwidth}
\fbox{\includegraphics[width=3in]{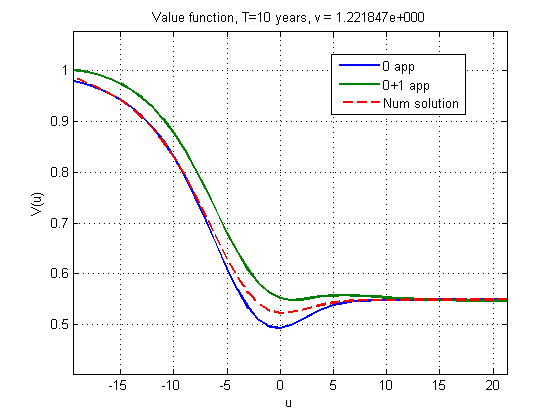}}
\end{minipage}
\hspace{0.1\textwidth}
\begin{minipage}[b]{0.4\textwidth}
\fbox{\includegraphics[width=3in]{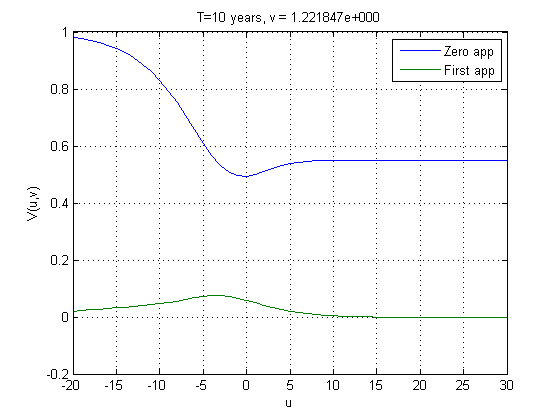}}
\end{minipage}
\caption{Value function $V(u,v)$ at $T$=10 yrs obtained by FD scheme, '0' and '0+1'
approximations. The initial parameters are given in Table~\ref{Tab}, test 1.}
\label{Comp1}
\end{figure}

\begin{figure}
\begin{minipage}[b]{0.4\textwidth}
\fbox{\includegraphics[width=3in]{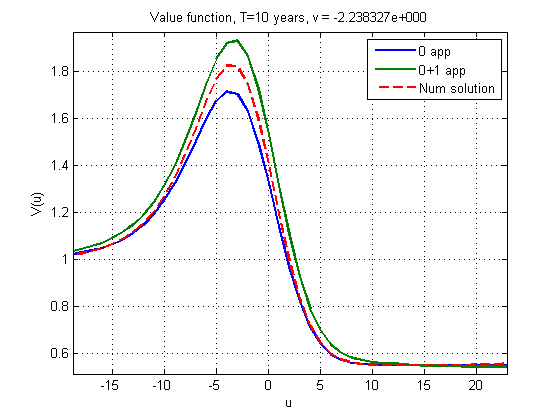}}
\end{minipage}
\hspace{0.1\textwidth}
\begin{minipage}[b]{0.4\textwidth}
\fbox{\includegraphics[width=3in]{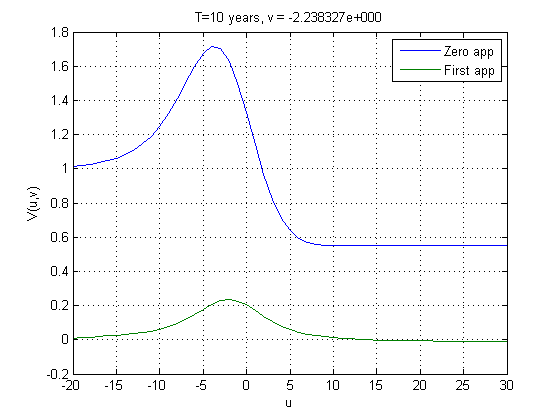}}
\end{minipage}
\caption{Value function $V(u,v)$ at $T$=10 yrs, comparison of the zero and first order approximations. The initial parameters are given in Table~\ref{Tab}, test 1.}
\label{Comp2}
\end{figure}

\begin{figure}
\begin{minipage}[b]{0.4\textwidth}
\fbox{\includegraphics[width=3in]{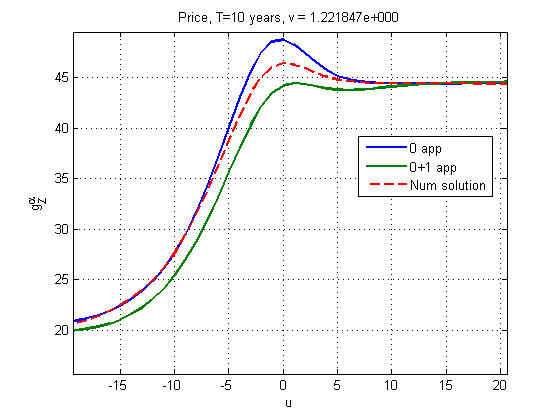}}
\end{minipage}
\hspace{0.1\textwidth}
\begin{minipage}[b]{0.4\textwidth}
\fbox{\includegraphics[width=3in]{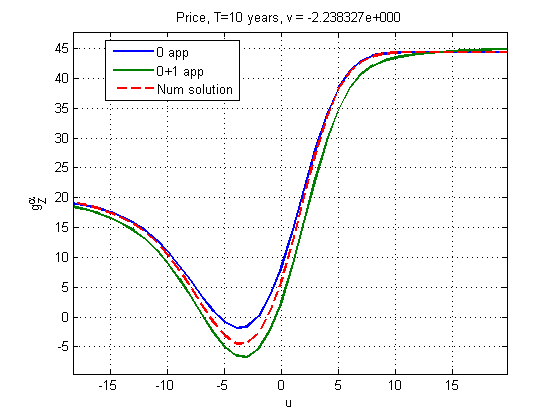}}
\end{minipage}
\caption{Price $g_Z^\alpha$ obtained in tests 1,2.}
\label{Price}
\end{figure}

In Fig~\ref{Comp1} the same quantities are computed for $v_0$ = 1.22 and $T=$10 yrs. Here the first plot presents comparison of the numerical solution with zero and '0+1' approximations. The second plot compares the zero and first order approximation. It is seen that the first approximation makes a small correction to the zero one in the region closed to $u=0$. Also both '0" and '0+1' approximations fit the exact numerical solution relatively well. This proves that our asymptotic closed form solution is robust. Results obtained with a second set of parameters (Test 2 in the Table~\ref{Tab}) are shown in Fig~\ref{Comp2}. In Fig~\ref{Price} we present price $g_Z^\alpha$ computed in tests 1,2 as a function of $ u $, where $p_Y$ is Black-Scholes put price with parameters of the corresponding tests.

\begin{figure}
\begin{minipage}[b]{0.4\textwidth}
\fbox{\includegraphics[width=3in]{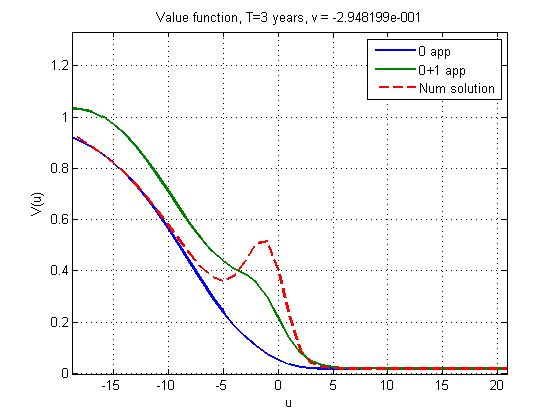}}
\end{minipage}
\hspace{0.1\textwidth}
\begin{minipage}[b]{0.4\textwidth}
\fbox{\includegraphics[width=3in]{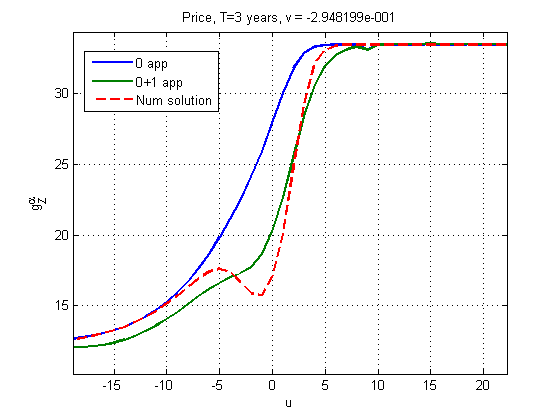}}
\end{minipage}
\caption{Value function $V(u,v)$ and price $g_Z^\alpha$ at $T$=3 yrs, comparison of the numerical solution,  '0' and '0+1' approximations at $\gamma = 0.2$. The initial parameters are given in Table~\ref{Tab}, test 1.}
\label{CompGamma3}
\end{figure}

\begin{figure}
\begin{minipage}[b]{0.4\textwidth}
\fbox{\includegraphics[width=3in]{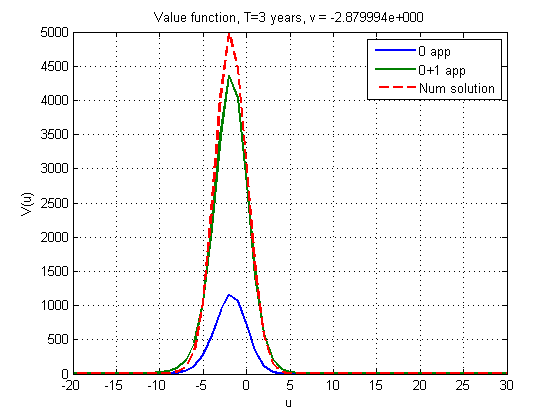}}
\end{minipage}
\hspace{0.1\textwidth}
\begin{minipage}[b]{0.4\textwidth}
\fbox{\includegraphics[width=3in]{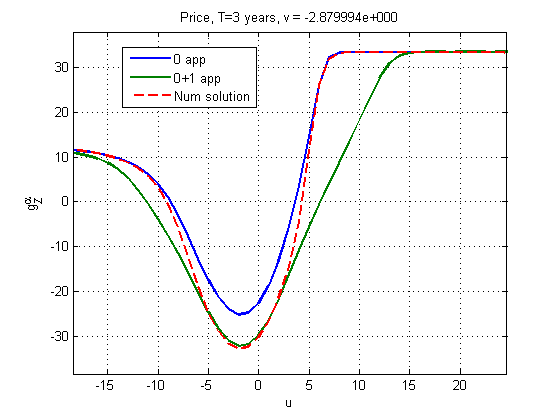}}
\end{minipage}
\caption{Value function $V(u,v)$ and price $g_Z^\alpha$ at $T$=3 yrs, comparison of the numerical solution,  '0' and '0+1' approximations at $\gamma = 0.2$. The initial parameters are given in Table~\ref{Tab}, test 1.}
\label{CompGamma4}
\end{figure}

Note that for  values of parameters used above the nonlinear term is small, therefore the solution is closed to the solution of the linear 2D heat equation obtained from the \eqref{Phi_uv_2} by omitting the nonlinear term. That is because $\gamma$ is small in our Test 1. To make testing more interesting, we changed $\gamma$ to $\gamma = 0.2$. Results obtained for $T$ = 3 yrs are presented in Fig~\ref{CompGamma3} for Test 1. and in Fig~\ref{CompGamma4} for Test 2. It is seen that in this case the '0+1' analytical approximation still fits the numerical solution.

The computational time in these tests is higher because the matrix root solver converges slower. The typical time at $T$=3yrs and $\alpha$ = 1 is 24 secs. Computation of the analytical approximation requires the same time as before which is about 1 sec.

If the exponent of the payoff function is positive, e.g. at $\alpha$ = 2 and other parameters as in Test 1, then the solution looks like a delta function. Under such conditions any numerical method experiences a problem being unable to resolve very high gradients within just few nodes. Therefore, in this case one has to exploit a non-uniform grid saturated close to the peak of the value function. This brings extra complexity to the numerical scheme while our analytical approximation is free of such problems.

On the other hand it is natural to statically hedge $C_Z$ option with the option $C_Y$ with same or close moneyness. This significantly reduces the exponent and allows one  to use the same mesh even at high values of the
risk aversion parameter $\gamma$.

As the numerical performance of the model depends on the value of $ \gamma $,
a few comments are due at this point. Though we do not address calibration
of our model in the present paper (leaving it for future work), the value
of $ \gamma $ should be found by calibrating the model to market data.
It is not entirely obvious how to do this since we use an illiquid asset $Z$, and
moreover the price of our
complex option $g_Z^\alpha$ is not an additive sum of its components
for incomplete markets. It therefore makes sense
to calibrate the model to another set of instruments that are both
liquid and strongly correlated with the original instruments.
In the context of equity options, such calibration of $ \gamma $
was done in \cite{ImpliedRiskAversion2010}, giving rise to
values of $ \gamma \sim 0.1-0.6 $. While it is not
exactly clear how $\gamma$ found in this setting is
related to $\gamma$ of our original problem, we expect the latter to be
of the same order of magnitude.

Note that the results in Fig.\ref{CompGamma4} clearly demonstrate that the proposed method is just asymptotic. Indeed, the first correction to the solution in Fig.\ref{CompGamma4} is a few times larger then the zero-order solution. As usual, there exists an optimal number of terms in the asymptotic series that fit
the exact solution best. We do not pursue such analysis in this paper\footnote{From theory we know that the asymptotic solution converges to the true solution
as $\varepsilon \rightarrow 0$, however due to our definition of $\varepsilon$ via $\rho_{yz}$ this is not an interesting case.}.

To characterize sharpness of the peak in the value function it is further convenient to introduce a new parameter $\psi = \gamma K_z$. If $\psi$ is high, e.g. $\psi > 50$ the value function is almost a delta function, so the asymptotic solution as well as its numerical counterpart are not expected to produce correct results
unless they are further modified. Based on the
results of \cite{ImpliedRiskAversion2010} one can see that the
value of $\psi$ calibrated to the market varies from 0.01 to 15. Therefore, in
our test 2 we used $\psi$ = 20. Both our numerical and
asymptotic methods work with no problem for these values of $ \psi $.

\begin{figure}
\begin{minipage}[b]{0.4\textwidth}
\fbox{\includegraphics[width=3in]{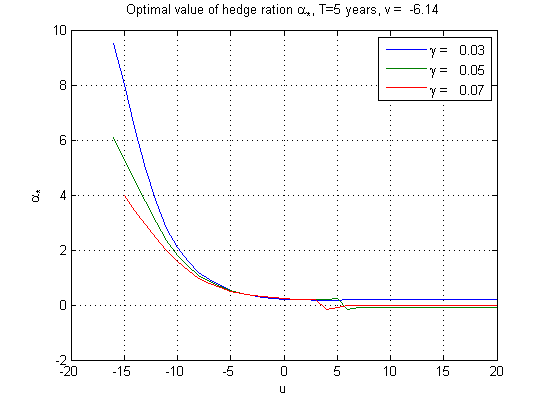}}
\end{minipage}
\hspace{0.1\textwidth}
\begin{minipage}[b]{0.4\textwidth}
\fbox{\includegraphics[width=2.95in]{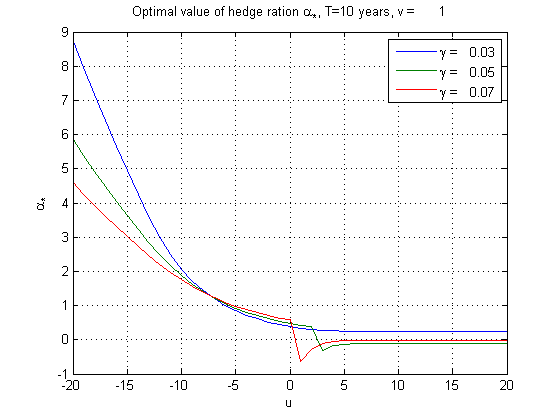}}
\end{minipage}
\caption{Optimal hedge $\alpha_*$ computed based on \eqref{indif_price_2} using the Brent method.}
\label{AlphaStar}
\end{figure}

In Fig~\ref{AlphaStar} the optimal hedge $\alpha_*$ is computed based on
\eqref{indif_price_2} which was solved using
Brent's method (\cite{SallehZomayaBakar2007}). The initial
parameters correspond to test 1 in Table~\ref{Tab} in the first plot, and
test 2 in the second plot. Note that for $u <-15$ (the first plot)
and $u < -20$ (the second plot), \eqref{indif_price_2} does not have a
minimum, so the maximum is obtained at the edge of the chosen
interval of $\alpha$. The latter could be defined based on some other
preferences of the trader, for instance, the total capital she wants to
invest into this strategy etc.

\begin{figure}
\begin{minipage}[b]{0.4\textwidth}
\fbox{\includegraphics[width=3in]{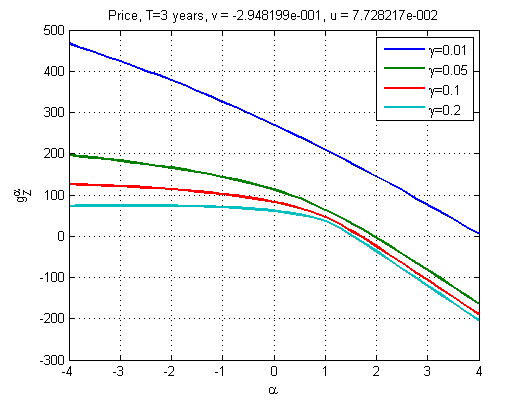}}
\end{minipage}
\hspace{0.1\textwidth}
\begin{minipage}[b]{0.4\textwidth}
\fbox{\includegraphics[width=3.05in]{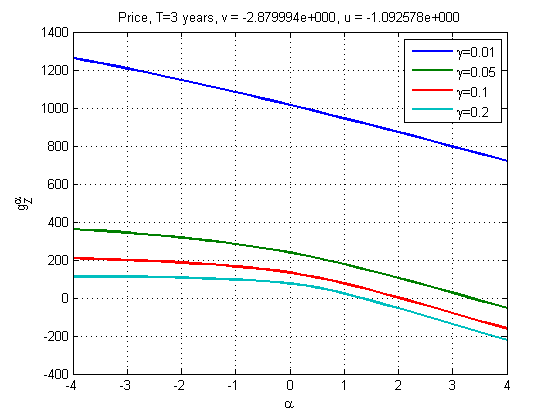}}
\end{minipage}
\caption{Price $g_Z^\alpha$ as a function of $\alpha$ at various $\gamma$ computed for the initial data in tests 1 and 2.}
\label{PriceAlphaGamma}
\end{figure}

Finally, in Fig~\ref{PriceAlphaGamma} price $g_Z^\alpha$ is presented as a
function of $\alpha$ for various $\gamma$.
It is seen that this function is convex which was first
showed in \cite{IlhanSircar2006} in a different setting. Note that
these results were obtained using a new numerical method
mentioned above. It combines Strang's splitting with the Fast Gaussian Transform,
and accelerates calculations approximately by factor 40 as
compared with a non-linear version of the 2d Crank-Nicholson scheme. A detailed
description of the method will be given elsewhere.

\section*{Acknowledgments}
We thank Peter Carr and  attendees of the "Global Derivatives USA 2011" conference for useful comments. I.H. would like to thank Andrew Abrahams and Julia Chislenko for support and interest in this work. We assume full responsibility for any remaining errors.

\newpage
\newcommand{\noopsort}[1]{} \newcommand{\printfirst}[2]{#1}
  \newcommand{\singleletter}[1]{#1} \newcommand{\switchargs}[2]{#2#1}

\newpage
\begin{appendices}
\markboth{Appendices}{Appendices}

\section{Asymptotic solutions of the \eqref{Phi_uv_2}} \label{A}

Here we describe in more detail the solutions for the zero and first order approximation in $\varepsilon$, and outline a generalization of our approach to an arbitrary order in $\varepsilon$.

\subsection{Zero-order approximation}
In the zero order approximation $\varepsilon = 0$ the \eqref{Phi_uv_2} does
not contain derivatives wrt $v$:
\begin{equation}
\label{zeroEq2}
\Phi_{0,\tau} = \frac{1}{2} \Phi_{0,uu} - \frac{1}{2} {\rho}_{xz}^2 \frac{ \left( \Phi_{0,u} \right)^2} {\Phi_0}
\end{equation}
The solution of this equation proceeds along similar lines to
Sect.~\ref{sect:zeroorder} using a change of dependent variable
\begin{equation} \label{distortion2}
\Phi_0(\tau,u,v) = \left[ \phi(\tau,u,v) \right]^{1/(1-{\rho}_{xz}^2)}
\end{equation}
\noindent which reduces \eqref{zeroEq2} to the heat equation
\begin{equation} \label{linPDE2}
\phi_{\tau} = \frac{1}{2} \phi_{uu},
\end{equation}
\noindent subject to the initial condition $\phi_{u,v,0} = \Phi(u,v,0)^{1-{\rho}_{xz}^2}$. The explicit form for the latter coincides with (\ref{init_cond_quand2}) provided we substitute $\bar{\gamma} = \gamma (1 - {\rho}_{xz}^2)$,  $\omega_1 = 1$, $\zeta = 1/\beta$ and $ \omega_2 = 0 $.

The explicit zero-order solution thus reads (compare with \eqref{Phi_0_sol})
\begin{equation} \label{Phi_0_sol2}
\Phi_0(u,v,\tau) =
\left[\frac{1}{\sqrt{2 \pi \tau}}\int_{-\infty}^{\infty} e^{-\frac{(u' - u)^2}{2 \tau}} \phi(u',v,0) d u'\right]^{1/(1-{\rho}_{xz}^2)}
\end{equation}
Note that \eqref{Phi_0_sol2} provides the general zero-order solution for the HJB equation with arbitrary initial conditions at $ \tau = 0 $. For our specific initial conditions \eqref{init_cond_quand2}, the solution is readily obtained in closed form in terms of the error (or normal cdf)  function (see Appendix \ref{A1}).

\subsection{First-order approximation}
For the first correction in the \eqref{Phi_uv_2} we obtain the following PDE
\begin{equation} \label{first}
\Phi_{1,\tau} = \frac{1}{2} \Phi_{1,uu} - \rho_{xz}^2 \frac{\Phi_{0,u}}{\Phi_0} \Phi_{1,u}
+ \frac{1}{2} \rho_{xz}^2 \left( \frac{\Phi_{0,u}}{\Phi_0} \right)^2 \Phi_{1} + \Theta_1(u,v,\tau)
\end{equation}
\noindent where $\Theta_1(u,v, \tau) = \theta_3 \Phi_{0,uv}$.

This equation coincides with \eqref{first2} except that the free term is different,
and the correlation parameter is $ \rho_{xz} $ rather than $ \bar{\rho}_{xz} $.
The solution proceeds as in Sect.~\ref{sect:Firstorder}, resulting in the following
expression:
\begin{equation} \label{FirstOrderSol1}
\Phi_1(u,v,\tau) = \Phi_0^{\rho_{xz}^2} (u,v,\tau)\int_0^\tau d\chi \int_{-\infty}^\infty
d u' \frac{e^{-\frac{(u- u' )^2}{2 (\tau-\chi)}} }{\sqrt{2 \pi (\tau-\chi)}} \Theta_1(u',v,\chi) \Phi_0^{-\rho_{xz}^2}(u',v,\chi) \\
\end{equation}
The double integral that enters this expression can be split out in two parts. One of them could be found in closed form, while the other one requires numerical computation (see Appendix \ref{A2}). Our numerical tests show that for many sets of the initial parameters the first integral is much higher than the second one, so the latter could be neglected. However, we were not able to identify in advance at which particular values of the parameters this could be done. Moreover, for some other initial parameters we observe an opposite situation.

\subsection{Second order approximation and higher orders}
The second order equation has the same form as \eqref{first}
\begin{equation*}
\Phi_{2,\tau} = \frac{1}{2} \Phi_{2,uu} - \rho_{xz}^2 \frac{\Phi_{0,u}}{\Phi_0} \Phi_{2,u}
+ \frac{1}{2} \rho_{xz}^2 \left( \frac{\Phi_{0,u}}{\Phi_0} \right)^2 \Phi_{2} + \Theta_2(u,v,\tau)
\end{equation*}
\noindent where
\begin{equation*}
\Theta_2(u,v, \tau) = \frac{1}{2}\theta_2 \Phi_{0,vv} + \theta_3 \left(\Phi_{1,uv} - \theta_3 \Phi_{0,uv} \right) - \frac{1}{2}\rho_{xz}^2 \Phi_0 \left(\Phi_1/\Phi_0\right)^{'2}_u
\end{equation*}
As this equation has to be solved also subject to zero initial conditions, the solution is obtained in the same way as above:
\begin{equation*}
\Phi_2(u,v,\tau) = \Phi_0^{\rho_{xz}^2}(u',v,\tau) \int_0^{\tau}\int_{-\infty}^\infty  \frac{e^{-\frac{(u-u' )^2}{2 (\tau-\chi)}} }{\sqrt{2 \pi (\tau-\chi)}} \Theta_2(u,v,\chi) \Phi_0^{-\rho_{xz}^2}(u',v,\chi) d\chi d u'
\end{equation*}

This shows that in higher order approximations in $\varepsilon$ the type of the equation to solve doesn't change as well as the initial conditions. Therefore, the solution to the $n$-th order approximation reads
\begin{equation*}
\Phi_n(u,v,\tau) =  \Phi_0^{\rho_{xz}^2}(u',v,\tau) \int_0^{\tau}\int_{-\infty}^\infty \frac{e^{-\frac{(u-u' )^2}{2 (\tau-\chi)}} }{\sqrt{2 \pi (\tau-\chi)}} \Theta_n(u,v,\chi) \Phi_0^{-\rho_{xz}^2}(u',v,\chi) d\chi d u'
\end{equation*}
\noindent where $\Theta_n$ can be expressed via already found solutions of order $i, i=1...n-1$ and their derivatives on $u$ and $v$. The exact representation for $\Theta_n$ follows combinatorial rules and reads
\begin{align}
\Theta_n(u,v,\tau) &= \theta_3 \sum _{i=1}^n \frac{1}{(n-i+1)!} \nfp{n-i+1}{\xi(\varepsilon)}{\varepsilon}\Bigg|_{\varepsilon=0} \Phi_{i-1, uv} \\
&+ \frac{1}{2}\theta_2 \sum _{i=2}^n \frac{1}{(n-i+2)!} \nfp{n-i+2}{\xi^2(\varepsilon)}{\varepsilon}\Bigg|_{\varepsilon=0} \Phi_{i-2, vv} \nonumber - \frac{1}{2} \rho_{xz}^2 \Xi_n, \nonumber
\end{align}
\noindent where $\Xi_n$ is a coefficient at $\varepsilon^{n-1}, \ n > 1$ in the following expansion
\begin{align*}
& \Phi_0 \left(\fp{\ln \Phi_0}{u}\right)^2 (1+\beta_3)\left[1 + \fp{\ln (1+\beta)}{u} \left(\fp{\ln \Phi_0}{u}\right)^{-1} \right]^2, \qquad \beta_3 = \sum_{i=1}^\infty \varepsilon^i \Phi_i/\Phi_0 \nonumber
\end{align*}
In particular, $\Xi_3$ reads
\begin{equation*}
\Xi_3 = \frac{\Phi_2 \Phi_{0,u}^2 + \Phi_{0} \left(\Phi_1^2 - 2\Phi_2 \Phi_{0,u}\right)}
{\Phi_0^2} - \frac{2 \Phi_1 \Phi_{1,u}}{\Phi_{0,u}} +\frac{\Phi_0 \Phi_{1,u}^2}{\Phi_{0,u}^2}
+2 \Phi_{2,u}
\end{equation*}
The explicit representation of the solutions of an arbitrary order in quadratures is important because per our definition of $\varepsilon$ the convergence of the \eqref{expan} is expected to be slow. Indeed, if one wants the final precision to be about 0.1 at $\rho_{yz} = 0.8$, the number of important terms $m$ in the expansion \eqref{expan} could be rawly calculated as $\varepsilon^m = 0.1$, which gives $m=4.5$.

All the integrals with $n > 1$ do not admit a closed form representation and have to be computed numerically.

\section{Closed form solutions for the zero-order approximation} \label{A1}
Since our payoff is a piece-wise function, the integral in \eqref{Phi_0_sol} can
be represented as a sum of three integrals. We denote
$\omega = \rho_{xy} v/\rho_{xz}$ and represent the zero-order solution as
follows:
\begin{equation*}
\Phi_0(u,v,\tau) = \left[J_1^{(\zeta)} + J_2^{(\zeta)} + J_3^{(\zeta)}\right]^{\frac{1}{1-\rho_{xz}^2}},
\qquad \zeta = {\rm sign}(v) \\
\end{equation*}
\noindent where sign $(-)$ means that $\omega = \rho_{xy} v/\rho_{xz} < 0$, and
sign $(+)$ - that $\omega > 0$, and
\begin{align} \label{Jdef}
J_1^{(+)} &= \frac{1}{\sqrt{2 \pi \tau}}\int_{- \infty}^{-\omega}d u' e^{-\frac{(u' - u)^2}{2 \tau}}
\exp\left[ - \bar{\gamma} \left( K_z e^{\sigma_z\frac{\rho_{xz}}{\rho_{xy}}  ( \omega + u')} - \alpha K_y e^{\sigma_y u'} \right) \right] \\
&= \Omega\left(-\omega, - \bar{\gamma} K_z e^{\sigma_z\frac{\rho_{xz}}{\rho_{xy}}\omega}, \sigma_z \frac{\rho_{xz}}{\rho_{xy}}, \bar{\gamma} \alpha K_y, \sigma_y\right), \nonumber \\
J_2^{(+)} &= \frac{1}{\sqrt{2 \pi \tau}}\int_{- \omega}^{0}d u' e^{-\frac{(u' - u)^2}{2 \tau}}
\exp\left[ - \bar{\gamma} \left(K_z - \alpha K_y e^{\sigma_y u'}\right) \right], \nonumber \\
&= \Omega\left(0, - \bar{\gamma} K_z, 0, \bar{\gamma} \alpha K_y, \sigma_y\right)
- \Omega\left(-\omega, - \bar{\gamma} K_z, 0, \bar{\gamma} \alpha K_y, \sigma_y\right), \nonumber \\
J_3^{(+)} &= \frac{e^{- \bar{\gamma} \left(K_z - \alpha K_y\right)}}{\sqrt{2 \pi \tau}}\int_{0}^{\infty}d u' e^{-\frac{(u' - u)^2}{2 \tau}} = e^{- \bar{\gamma} \left(K_z - \alpha K_y\right)} \left[1 -\frac{1}{2} {\rm Erfc}\left(\frac{u}{\sqrt{2\tau}}\right)\right], \nonumber  \\
J_1^{(-)} &= \frac{1}{\sqrt{2 \pi \tau}}\int_{- \infty}^{0}d u' e^{-\frac{(u' - u)^2}{2 \tau}}
\exp \left[ - \bar{\gamma} \left(K_z e^{\sigma_z \frac{\rho_{xz}}{\rho_{xy}} (\omega + u')}
- \alpha K_y e^{\sigma_y u'} \right) \right] \nonumber \\
&= \Omega\left(0, - \bar{\gamma} K_z e^{\sigma_z\frac{\rho_{xz}}{\rho_{xy}}\omega},
\sigma_z \frac{\rho_{xz}}{\rho_{xy}},
\bar{\gamma} \alpha K_y, \sigma_y\right), \nonumber \\
J_2^{(-)} &= \frac{1}{\sqrt{2 \pi \tau}}\int_{0}^{-\omega}d u' e^{-\frac{(u' - u)^2}{2 \tau}} \exp\left[ - \bar{\gamma} \left( K_z e^{\sigma_z \frac{\rho_{xz}}{\rho_{xy}}(\omega + u')}
- \alpha K_y \right) \right] \nonumber \\
&= \Omega\left(-\omega, - \bar{\gamma} K_z e^{\sigma_z\frac{\rho_{xz}}{\rho_{xy}}\omega},
\sigma_z \frac{\rho_{xz}}{\rho_{xy}}, \bar{\gamma} \alpha K_y, 0\right)
- \Omega\left(0, - \bar{\gamma} K_z e^{\sigma_z\frac{\rho_{xz}}{\rho_{xy}}\omega},
\sigma_z \frac{\rho_{xz}}{\rho_{xy}}, \bar{\gamma} \alpha K_y, 0\right), \nonumber \\
J_3^{(-)} &= \frac{e^{- \bar{\gamma} \left(K_z - \alpha K_y \right)}}{\sqrt{2 \pi \tau}}\int_{-\omega}^{\infty}d u'  e^{-\frac{(u' - u)^2}{2 \tau}} = \frac{1}{2} e^{- \bar{\gamma} \left(K_z - \alpha K_y \right)} {\rm Erfc}\left(-\frac{u+\omega}{\sqrt{2\tau}}\right). \nonumber
\end{align}
Here ${\rm Erfc}(x)$ is the complementary error function, and
\begin{align} \label{Omega}
\Omega(a,\delta,p,\beta,q) &\equiv \frac{1}{\sqrt{2 \pi \tau}} \int_{-\infty}^a e^{\delta e^{p u'} + \beta e^{q u'}}e^{-\frac{(u' - u)^2}{2 \tau}} du' \\
&= \frac{1}{\sqrt{2 \pi \tau}}\sum _{i=0}^\infty \sum _{j=0}^i \frac{\delta^{i-j} \beta^j}{j! (i-j)!}\int_{-\infty}^a e^{[p(i-j) + q j] u'} e^{-\frac{(u' - u)^2}{2 \tau}} du' \nonumber \\
&= \frac{1}{2}\sum _{i=0}^\infty \sum _{j=0}^i \frac{\delta^{i-j} \beta^j}{j! (i-j)!}
e^{A\left(u+A \frac{\tau}{2}\right)}{\rm Erfc}\left(\frac{u - a + A \tau}{\sqrt{2\tau}} \right), \nonumber \\
A &= i p + j (q - p). \nonumber
\end{align}
Since the complementary error function quickly approaches zero with $x \ll 0$, or 1 with $x \gg 0$, the number of terms one has to keep in the above sums should not be high.

We also need the following function
\begin{equation} \label{calJ}
\mathcal{J}_{u,v} = \frac{1}{\sqrt{2 \pi \tau}}\int_{-\infty}^{\infty} e^{-\frac{(u' - u)^2}{2 \tau}} \phi_{u',v}(u',v,0) d u',
\end{equation}
\noindent when computing the first order approximation. Using a general form of the payoff
$\phi(u,v,0) = e^{\delta(v) e^{p u}+ \beta e^{q u}}$ \footnote{Compare with \eqref{init_cond_quand2}.}, it can be represented in the form
\begin{align} \label{Jouv}
\mathcal{J}_{u,v} &= \frac{1}{\sqrt{2 \pi \tau}}\int_{-\infty}^{\infty} e^{-\frac{(u' - u)^2}{2 \tau}}  \phi_{u',v}(u',v,0) d u' \\
&= \delta'(v)\frac{1}{\sqrt{2 \pi \tau}}\int_{-\infty}^{\infty} e^{-\frac{(u' - u)^2}{2 \tau}} e^{p u'} \phi(u',v,0) \left(p+e^{q u'} q \beta +e^{p u'} p \delta(v)\right)d u' \nonumber \\
&= \mathcal{J}_{1,u,v}^{(\zeta)} + \mathcal{J}_{2,u,v}^{(\zeta)} + \mathcal{J}_{3,u,v}^{(\zeta)}, \nonumber
\end{align}
\noindent where
\begin{align*}
\mathcal{J}_{2,u,v}^{(+)} &= \mathcal{J}_{3,u,v}^{(+)} = \mathcal{J}_{3,u,v}^{(-)} = 0, \\
\mathcal{J}_{1,u,v}^{(+)} &= - \bar{\gamma} K_z \sigma_z e^{\sigma_z\frac{\rho_{xz}}{\rho_{xy}} \omega}
\Omega_1\left(-\omega, - \bar{\gamma} K_z e^{\sigma_z\frac{\rho_{xz}}{\rho_{xy}}\omega}, \sigma_z \frac{\rho_{xz}}{\rho_{xy}}, \bar{\gamma} \alpha K_y, \sigma_y\right)
\nonumber \\
\mathcal{J}_{1,u,v}^{(-)} &= - \bar{\gamma} K_z \sigma_z e^{\sigma_z\frac{\rho_{xz}}{\rho_{xy}} \omega} \Omega_1\left(0, - \bar{\gamma} K_z e^{\sigma_z\frac{\rho_{xz}}{\rho_{xy}}\omega},
\sigma_z \frac{\rho_{xz}}{\rho_{xy}}, \bar{\gamma} \alpha K_y, \sigma_y\right), \nonumber \\
\mathcal{J}_{2,u,v}^{(-)} &= - \bar{\gamma} K_z \sigma_z e^{\sigma_z\frac{\rho_{xz}}{\rho_{xy}} \omega} \Bigg\{\Omega_1\left(-\omega, - \bar{\gamma} K_z e^{\sigma_z\frac{\rho_{xz}}{\rho_{xy}}\omega},
\sigma_z \frac{\rho_{xz}}{\rho_{xy}}, \bar{\gamma} \alpha K_y, 0\right) \nonumber \\
&- \Omega_1\left(0, - \bar{\gamma} K_z e^{\sigma_z\frac{\rho_{xz}}{\rho_{xy}}\omega},
\sigma_z \frac{\rho_{xz}}{\rho_{xy}}, \bar{\gamma} \alpha K_y, 0\right) \Bigg\}, \nonumber
\end{align*}
\noindent and
\begin{align} \label{Omega1}
\Omega_1(a,\delta,p,\beta,q) &= \frac{1}{2}\sum _{i=0}^\infty \sum _{j=0}^i \frac{\delta^{i-j} \beta^j}{j! (i-j)!}
\left[ p \Lambda(p) + p \delta \Lambda(2p) + q \beta \Lambda(p+q)\right], \\
\Lambda(b) &\equiv e^{A(b)\left[u+A(b) \frac{\tau}{2}\right]}{\rm Erfc}\left(\frac{u - a + A(b) \tau}{\sqrt{2\tau}} \right), \qquad A(b) \equiv i p + j (q - p) + b. \nonumber
\end{align}

\section{Transformation of the first order solution of the \eqref{Phi_uv_2}} \label{A2}
The first order approximation is given by \eqref{FirstOrderSol} where
the zero-order solution $\Phi_0(u,v,\tau)$ has been already
computed in Appendix~\ref{A1}. We plug in this solution into
\eqref{FirstOrderSol} to obtain
\begin{align} \label{Int1App}
\Phi_1&(u,v,\tau) = \Phi_0^{\rho_{xz}^2} (u,v,\tau)\int_0^\tau d\chi \int_{-\infty}^\infty
d u' \frac{e^{-\frac{(u- u' )^2}{2 (\tau-\chi)}} }{\sqrt{2 \pi (\tau-\chi)}} \Theta_1(u',v,\chi) \Phi_0^{-\rho_{xz}^2}(u',v,\chi) \\
&= \theta_3 \Phi_0^{\rho_{xz}^2} (u,v,\tau)\int_0^\tau d\chi \int_{-\infty}^\infty
d u' \frac{e^{-\frac{(u- u' )^2}{2 (\tau-\chi)}} }{\sqrt{2 \pi (\tau-\chi)}}
J(u',v,\chi)^{\frac{-\rho^2_{xz}}{1-\rho^2_{xz}}} \partial_{u',v} J(u',v,\chi)^{\frac{1}{1-\rho^2_{xz}}},  \nonumber \\
J &= \left[J_1^{(\zeta)} + J_2^{(\zeta)} + J_3^{(\zeta)}\right] \nonumber
\end{align}
\noindent where the integrals $J_i^{(\zeta)}, \ i=1,3$ are defined in \eqref{Jdef}.

The internal integral can be simplified. Indeed, since
\begin{equation*}
J^{\frac{-\rho^2_{xz}}{1-\rho^2_{xz}}} \partial_{u,v} J^{\frac{1}{1-\rho^2_{xz}}} =
\frac{1}{1-\rho^2_{xz}}J_{u',v} + \frac{\rho^2_{xz}}{(1-\rho^2_{xz})^2} \frac{J_v J_{u'}}{J}
\end{equation*}
\noindent the internal integral in \eqref{Int1App} can be represented as a
sum of two integrals $\mathfrak{I}_1 + \mathfrak{I}_2$, where
\begin{align*}
\mathfrak{I}_1 &= \frac{1}{1-\rho^2_{xz}} \int_{-\infty}^\infty d u' \frac{e^{-\frac{(u- u' )^2}{2 (\tau-\chi)}} }{\sqrt{2 \pi (\tau-\chi)}} J_{u',v}, \\
\mathfrak{I}_2 &= \frac{\rho^2_{xz}}{(1-\rho^2_{xz})^2} \int_{-\infty}^\infty d u' \frac{e^{-\frac{(u- u' )^2}{2 (\tau-\chi)}} }{\sqrt{2 \pi (\tau-\chi)}} \frac{J_v J_{u'}}{J}
\end{align*}

\begin{figure}
\begin{minipage}[b]{0.4\textwidth}
\fbox{\includegraphics[width=3in]{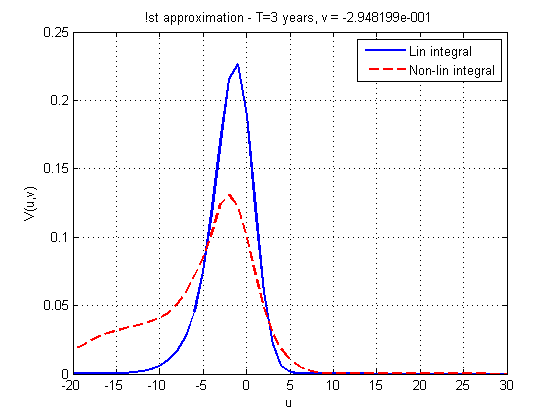}}
\end{minipage}
\hspace{0.1\textwidth}
\begin{minipage}[b]{0.4\textwidth}
\fbox{\includegraphics[width=3in]{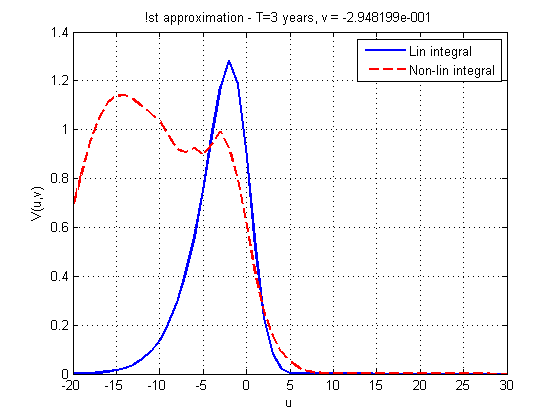}}
\end{minipage}
\caption{Comparison of $\mathfrak{I}_1$ and $\mathfrak{I}_2$ at $T$=3 yr and $\gamma$ = 0.03 and 0.2, test 1.}
\label{FigJ2}
\end{figure}

\begin{figure}
\begin{minipage}[b]{0.4\textwidth}
\fbox{\includegraphics[width=3in]{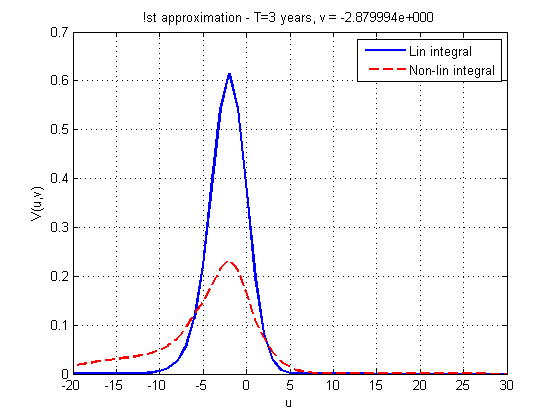}}
\end{minipage}
\hspace{0.1\textwidth}
\begin{minipage}[b]{0.4\textwidth}
\fbox{\includegraphics[width=3in]{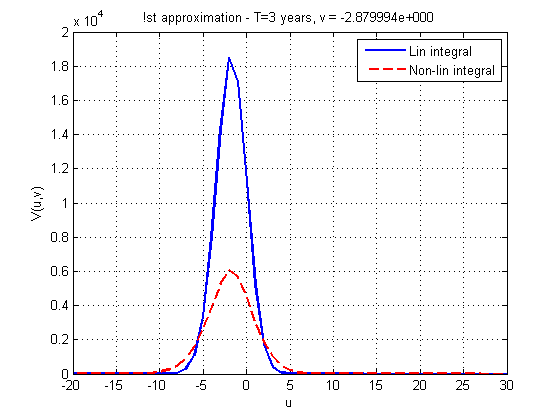}}
\end{minipage}
\caption{Comparison of $\mathfrak{I}_1$ and $\mathfrak{I}_2$ at $T$=3 yr and $\gamma$ = 0.03 and 0.2, test 2.}
\label{FigJ1}
\end{figure}

As we already mentioned the second integral could be either smaller or
larger than the
second one depending on the parameters.
This is illustrated in Fig.~\ref{FigJ2}. The initial parameters are
taken from test 1 in Table~\ref{Tab} with $\alpha$ = 1 and $T$ = 3 yrs.
In Fig.~\ref{FigJ1} the same calculation is shown for the
parameters corresponding to test 2 in Table~\ref{Tab}.

The first integral $\mathfrak{I}_1$ can be rewritten using the
integral representation of $J$ in \eqref{Phi_0_sol}
\begin{align*}
\mathfrak{I}_1 &= \frac{1}{1-\rho^2_{xz}} \int_{-\infty}^\infty d u' \frac{e^{-\frac{(u- u' )^2}{2 (\tau-\chi)}} }{\sqrt{2 \pi (\tau-\chi)}} \partial_{u'} \int_{-\infty}^\infty d u'' \frac{e^{-\frac{(u''- u' )^2}{2 \chi}} }{\sqrt{2 \pi \chi}} \phi_{0,v}(u'',v,0) \\
&= \frac{1}{1-\rho^2_{xz}} \int_{-\infty}^\infty d u'' \phi_{0,v}(u'',v,0) \int_{-\infty}^\infty d u' \frac{e^{-\frac{(u- u' )^2}{2 (\tau-\chi)}} }{\sqrt{2 \pi (\tau-\chi)}} \frac{u''-u'}{\chi} \frac{e^{-\frac{(u''- u' )^2}{2 \chi}}}{{\sqrt{2 \pi \chi}}}  \\
&= \frac{1}{1-\rho^2_{xz}} \int_{-\infty}^\infty d u'' \phi_{0,v}(u'',v,0)
\frac{ e^{-\frac{(u-u'')^2}{2 \tau }}}{\sqrt{2 \pi} \tau ^{3/2}}(u''-u)
\end{align*}
Substituting this into \eqref{Int1App} and integrating over $\chi$, we find
\begin{align}
\Phi_1(u,v,\tau) &=  \frac{\theta_3 }{1-\rho^2_{xz}} \Phi_0^{\rho_{xz}^2} (u,v,\tau)\int_0^\tau d\chi \int_{-\infty}^\infty
d u'' \phi_{0,v}(u'',v,0) \frac{ e^{-\frac{(u-u'')^2}{2 \tau }}}{\sqrt{2 \pi} \tau ^{3/2}}(u''-u) \\
&= \frac{\theta_3}{1-\rho^2_{xz}} \Phi_0^{\rho_{xz}^2} (u,v,\tau) \int_{-\infty}^\infty
d u'' \phi_{0,v}(u'',v,0) \frac{ e^{-\frac{(u-u'')^2}{2 \tau }}}{\sqrt{2 \pi \tau}} (u''-u) \nonumber \\
&= -\frac{\theta_3\tau}{1-\rho^2_{xz}} \Phi_0^{\rho_{xz}^2} (u,v,\tau) \mathcal{J}_{u,v}.
\end{align}
Thus, the first correction to the solution obtained in the zero order approximation on $\varepsilon$ is approximately
$-\theta_3 \tau \frac{\sqrt{1-\rho^2_{yz}}}{1-\rho^2_{xz}} \Phi_0^{\rho_{xz}^2} (u,v,\tau) \mathcal{J}_{u,v}$, and the full solution in the "0+1" approximation reads
\[
\Phi(u,v,\tau) = \Phi_0(u,v,\tau) -\theta_3 \tau \frac{\sqrt{1-\rho^2_{yz}}}{1-\rho^2_{xz}} \Phi_0^{\rho_{xz}^2} (u,v,\tau) \mathcal{J}_{u,v}
\]

\section{Transformation of the first order solution of \eqref{Phi_vv_2}}
\label{B2}
The first order approximation is given by \eqref{FirstOrderSol}. It is
assumed that the zero-order solution $\Phi_0(u,v,\tau)$ is already
computed by using either the method presented in Appendix~\ref{A1}, or
using the Fast Gauss Transform. We plug in this solution
into \eqref{FirstOrderSol} to obtain
\begin{align} \label{Int1App2}
\Phi_1&(u,v,\tau) = \Phi_0^{\bar{\rho}_{xz}^2} (u,v,\tau)\int_0^\tau d\chi \int_{-\infty}^\infty
d u' \frac{e^{-\frac{(u- u' )^2}{2 (\tau-\chi)}} }{\sqrt{2 \pi (\tau-\chi)}} \Theta_1(u',v,\chi) \Phi_0^{-\bar{\rho}_{xz}^2}(u',v,\chi) \\
&= \frac{1}{2}\theta_2 \Phi_0^{\bar{\rho}_{xz}^2} (u,v,\tau)
\int_0^\tau d\chi \int_{-\infty}^\infty
d u' \frac{e^{-\frac{(u- u' )^2}{2 (\tau-\chi)}} }{\sqrt{2 \pi (\tau-\chi)}}
J(u',v,\chi)^{\frac{-\bar{\rho}^2_{xz}}{1-\bar{\rho}^2_{xz}}}  \partial_{v,v} J(u',v,\chi)^{\frac{1}{1-\bar{\rho}^2_{xz}}},  \nonumber \\
J &= \left[J_1^{(\zeta)} + J_2^{(\zeta)} + J_3^{(\zeta)}\right] \nonumber
\end{align}
\noindent where the integrals $J_i^{(\zeta)}, \ i=1,3$ are defined in \eqref{Jdef}.

It can be seen that \eqref{Int1App2} is similar to \eqref{Int1App} if
one replaces $\rho_{xz}$ with $\bar{\rho}_{xz}$
and $\partial_{v,v} J(u',v,\chi)^{\frac{1}{1-\bar{\rho}^2_{xz}}}$
with $\partial_{v,v} J(v,v,\chi)^{\frac{1}{1-\bar{\rho}^2_{xz}}}$. Therefore, we
can use the same idea as in Appendix~\ref{A2} to further simplify this integral.

Accordingly the inner integral in \eqref{Int1App2} can be rewritten as
\begin{equation*}
J^{\frac{-\bar{\rho}^2_{xz}}{1-\bar{\rho}^2_{xz}}} \partial_{v,v} J^{\frac{1}{1-\bar{\rho}^2_{xz}}} =
\frac{1}{1-\bar{\rho}^2_{xz}}J_{v,v} + \frac{\bar{\rho}^2_{xz}}{(1-\bar{\rho}^2_{xz})^2} \frac{J_v^2}{J}
\end{equation*}
\noindent The internal integral in \eqref{Int1App2} can then
be represented as a sum of two integrals $\mathfrak{I}_1 + \mathfrak{I}_2$, where
\begin{align*}
\mathfrak{I}_1 &= \frac{1}{1-\bar{\rho}^2_{xz}} \int_{-\infty}^\infty d u' \frac{e^{-\frac{(u- u' )^2}{2 (\tau-\chi)}} }{\sqrt{2 \pi (\tau-\chi)}} J_{v,v}, \\
\mathfrak{I}_2 &= \frac{\bar{\rho}^2_{xz}}{(1-\bar{\rho}^2_{xz})^2} \int_{-\infty}^\infty d u' \frac{e^{-\frac{(u- u' )^2}{2 (\tau-\chi)}} }{\sqrt{2 \pi (\tau-\chi)}} \frac{J_v^2}{J}
\end{align*}

The first integral $\mathfrak{I}_1$ can be modified using the integral
representation of $J$ in \eqref{Phi_0_sol}
\begin{align*}
\mathfrak{I}_1 &= \frac{1}{1-\rho^2_{xz}} \int_{-\infty}^\infty d u' \frac{e^{-\frac{(u- u' )^2}{2 (\tau-\chi)}} }{\sqrt{2 \pi (\tau-\chi)}}  \int_{-\infty}^\infty d u'' \frac{e^{-\frac{(u''- u' )^2}{2 \chi}} }{\sqrt{2 \pi \chi}} \phi_{0,vv}(u'',v,0) \\
&= \frac{1}{1-\bar{\rho}^2_{xz}} \int_{-\infty}^\infty d u'' \phi_{0,vv}(u'',v,0) \int_{-\infty}^\infty d u' \frac{e^{-\frac{(u- u' )^2}{2 (\tau-\chi)}} }{\sqrt{2 \pi (\tau-\chi)}} \frac{e^{-\frac{(u''- u' )^2}{2 \chi}}}{{\sqrt{2 \pi \chi}}}  \\
&= \frac{1}{1-\bar{\rho}^2_{xz}} \int_{-\infty}^\infty d u'' \phi_{0,vv}(u'',v,0)
\frac{ e^{-\frac{(u-u'')^2}{2 \tau }}}{\sqrt{2 \pi \tau}}
\end{align*}
Substituting this into \eqref{Int1App2} and integrating over $\chi$, we find
\begin{align}
\Phi_1(u,v,\tau) &=  \frac{1}{2} \frac{\theta_2}{1-\bar{\rho}^2_{xz}} \Phi_0^{\bar{\rho}_{xz}^2} (u,v,\tau)\int_0^\tau d\chi \int_{-\infty}^\infty
d u'' \phi_{0,vv}(u'',v,0) \frac{ e^{-\frac{(u-u'')^2}{2 \tau }}}{\sqrt{2 \pi \tau}} \\
&= \frac{\tau}{2}\frac{\theta_2}{1-\bar{\rho}^2_{xz}} \Phi_0^{\bar{\rho}_{xz}^2} (u,v,\tau) \int_{-\infty}^\infty
d u'' \phi_{0,vv}(u'',v,0) \frac{ e^{-\frac{(u-u'')^2}{2 \tau }}}{\sqrt{2 \pi \tau}} \nonumber
\end{align}

Thus, the first correction to the solution obtained in the zero order approximation on $\mu$ is approximately
$\frac{1}{2} \tau \theta_2 \frac{\mu}{1-\bar{\rho}^2_{xz}} \Phi_0^{\bar{\rho}_{xz}^2} (u,v,\tau) \mathcal{J}_{v,v}$, and the full solution in the "0+1" approximation reads
\[
\Phi(u,v,\tau) = \Phi_0(u,v,\tau) + \frac{1}{2} \tau \theta_2 \frac{\mu}{1-\bar{\rho}^2_{xz}} \Phi_0^{\bar{\rho}_{xz}^2} (u,v,\tau) \mathcal{J}_{v,v}.
\]

\end{appendices}
\vspace{1in}
\end{document}